\documentclass[12pt]{article}

\usepackage{scicite}

\usepackage{times}

\usepackage{graphicx}
\usepackage{mathrsfs}    
\usepackage{bm}             
\usepackage{upgreek}        
\usepackage[scr=boondox,bb= pazo]{mathalfa}    
\usepackage{amsmath}
\usepackage{lineno}
\usepackage{caption}

\usepackage[dvipsnames]{xcolor} 

\newenvironment{sciabstract}{%
\begin{quote} \bf}
{\end{quote}}

\title{Initial Foundation for Predicting Individual Earthquake's Location and Magnitude by Using Glass-Box Physics Rule Learner}

\author
{In Ho Cho,$^{1\ast}$ 
\\
\normalsize{$^{1}$CCEE Department, Iowa State University, Ames, IA 50011, USA}\\
\\
\normalsize{$^\ast$To whom correspondence should be addressed; E-mail:  icho@iastate.edu.}
}

\date{}

\begin{document} 


\baselineskip24pt


\maketitle 

\captionsetup[figure]{labelfont={bf},labelformat={default},labelsep=period,name={Fig.}}


\begin{sciabstract}
Although researchers accumulated knowledge about seismogenesis and decades-long earthquake data, predicting imminent individual earthquakes at a specific time and location remains a long-standing enigma. This study hypothesizes that the observed data conceal the hidden rules which may be unraveled by a novel glass-box (as opposed to black-box) physics rule learner (GPRL) framework. Without any predefined earthquake-related mechanisms or statistical laws, GPRL's two essentials, convolved information index and transparent link function, seek generic expressions of rules directly from data. GPRL's training with 10-years data appears to identify plausible rules, suggesting a combination of the pseudo power and the pseudo vorticity of released energy in the lithosphere. Independent feasibility test supports the promising role of the unraveled rules in predicting earthquakes' magnitudes and their specific locations. The identified rules and GPRL are in their infancy requiring substantial improvement. Still, this study hints at the existence of the data-guided hidden pathway to imminent individual earthquake prediction.
\end{sciabstract}
\newpage 
Analytically or computationally simulated earthquakes are widely used to offer valuable insights into the long-standing enigma of seismogenesis. Researchers seek clues from basic physics -- the thermal instability for computational reproductions of deeper slow earthquakes \cite{Wang:2020}, natural fluid injections into fault zones for earthquake swarms \cite{Ross:2020}, or sliding frictional blocks for the chaotic slip pulse behaviors \cite{Elbanna:2012}. By combining a number of mechanics-/physics-based rules, researchers can reproduce ``virtual'' earthquakes on computer \cite{Jiang:2016, Barbot:2012}. For illustration purposes, this paper calls these methods as ``bottom-up'' approach since their common starting point is the adopted mechanics- or physics-based rules and the associated parameters. Despite their important roles and values, the bottom-up approaches may explain real earthquake behaviors from a restricted angle, bounded by the intrinsic limits of the adopted rules and experiments used for determining the rules' key parameters. For instance, many studies used the frictional strength of the fault, e.g., rate-and-state friction \cite{Allison:2018, Zhu:2020, Rice:2001}, $f(\psi,V)=a\text{sinh}^{-1}\left( \frac{V}{2V_0}e^{\psi/a} \right)$ along with a rule about state evolution \cite{Ruina:1983, Marone:1998}, $G(\psi,V)=b\frac{V_0}{d_c}\left( e^{(f_0-\psi)/b} - \frac{V}{V_0} \right)$, where $a$ is the direct effect parameter, $V$ is the slip velocity, $V_0$ is the reference velocity, $\psi$ is the state variable, $b$ is the state evolution parameter, $d_c$ is the state evolution distance, and $f_0$ is the reference friction coefficient for steady sliding at $V_0$. Parameters $(a,b)$ are useful to simulate the depth of earthquake arrest or nucleation as well as physically sound fault behaviors (e.g., $a-b>0$ for the stable sliding, the so-called velocity-strengthening whereas $a-b<0$ for unstable sliding, the velocity-weakening). To determine the parameters, researchers often assume a ``link'' between laboratory tests and real-world earthquakes, e.g., wet granite laboratory tests for deriving $(a,b)$ \cite{Allison:2018, Blanpied:1991, Blanpied:1995}. The assumed ``link'' behind the experiment-driven parameters is compelling since it is built upon actual, physical observations. However, it is also true that such a link essentially brings in a simplification of fitting or extrapolations \cite{Allison:2018} and sometimes needs to embrace discrepancy among experimental observations, e.g., valuable yet diverse pieces of evidence in \cite{Blanpied:1991, Blanpied:1995, Mitchell:2016}. This paper calls another approach of using high-precision observation methodologies as ``top-down'' approach. The top-down approach becomes important with the advent of new observation technologies, which provide a valuable top-down viewpoint to explaining the fault, fracture, and slip behaviors. For instance, interferometric synthetic aperture radar (InSAR) \cite{Xu:2020, Xu:2020_Science} can help researchers constraint detailed surface deformations of large continental earthquakes and also can determine the sense of fault slips through image analyses, revealing many small fractures adjacent to rupture zones. Also, a densely distributed global positioning system (GPS) can help describe high-precision coseismic displacements \cite{Simon:2011}. The statistical approach constitutes a backbone of seismogenesis research. Scientists have recorded and documented them offering valuable statistical rules. The statistical laws of earthquakes such as the Gutenberg-Richter frequency-magnitude relation \cite{Gutenberg:1954} shed light on the long-term earthquake forecast \cite{Rundle:2003} from precursory patterns \cite{Keilis-Borok:2003} or small-scale events \cite{Tiampo:2002}.
\newline
It is natural for earthquake forecasting methods to seek to integrate these top-down, bottom-up, and statistical approaches holistically. Existing earthquake forecasting methods often rely on statistical/probabilistic assumptions about earthquake rate models, geophysical knowledge of known faults' characteristics, physics assumptions like Coulomb stress-based criteria, and many other empirical models \cite{Toda:2020, Field:2015}. For instance, UCERF3 (the Third Uniform California Earthquake Rupture Forecast) \cite{Field:2015} inherited the elastic-rebound theory and helps incorporate other seismogenic mechanisms, e.g., UCERF3-ETAS \cite{Field:2017} -- an extension to the epidemic-type aftershock sequence (ETAS) of \cite{Ogata:1998}, thereby helping ``operational'' (providing an official protocol for real-time forecasts to the public) earthquake forecasting in California \cite{Milner:2020, Page:2014}. Combined with advanced statistics, ETAS-based forecasting offers important statistics about collective seismicity, holding practical and scientific importance \cite{Shcherbakov:2019, Nandan:2021}. However, these forecasting methods' accuracy in predicting ``individual'' earthquake's specific location is limited by the underlying statistical, empirical, and physical assumptions. It is important to note the central difference of the present work from existing forecasting methods: first, this paper focuses on predicting ``individual'' future earthquake's location and magnitude in lieu of collective event counts or overall probability; second, this paper intentionally uses the observed earthquake data without adopting any pre-defined statistical laws (e.g., various power laws like GR law, Omori law, fertility law, or magnitude-energy law), in hopes of unraveling hidden rules guided by data only. 
Recently, newly emerging technologies of machine learning (ML) gradually play an important role in earthquake-related research. For instance, deep learning is harnessed to study earthquake swarms \cite{Ross:2020} and to improve seismic phase-detection \cite{Ross:2018}. The convolutional networks are used for seismic phase picking by \cite{Pardo:2019}. Despite their notable contributions, ML-driven exploration of hidden mechanisms behind earthquakes is in its infancy. Limits of direct use of existing ML methods are summarized in \cite{Supplementary}.
\newline
The common challenge of the aforementioned approaches in pursuit of hidden rules behind imminent earthquake predictions is that they look at earthquakes through a pre-defined lens of scientists. Still, earthquake remains a chaotic, natural enigma involving the multifaceted hidden physics (Fig. \ref{fig:Multifaceted_Physics}). If one truly seeks to unravel a ``hidden rule'' itself, it may be vital to remain independent of pre-defined rules without any prejudice. Indeed, recent efforts appear to support that earthquake prediction is a feasible scientific question \cite{Varotsos:2020}. This study hypothesizes that the most reliable source would be the observed data which conceal the hidden rules. This study also hypothesizes that exploring and learning the observed earthquake data may help unravel hidden rules behind imminent earthquake prediction. This study places top priority on the minimal use of general physics and scientists' eyes independent of earthquake-related pre-defined mechanisms and lets GPRL identify a plausible and interpretable generic expression of the hidden rule of the imminent earthquake prediction.  
\section*{Results}
\begin{figure}
  \centering
  \includegraphics[width=1.0\textwidth]{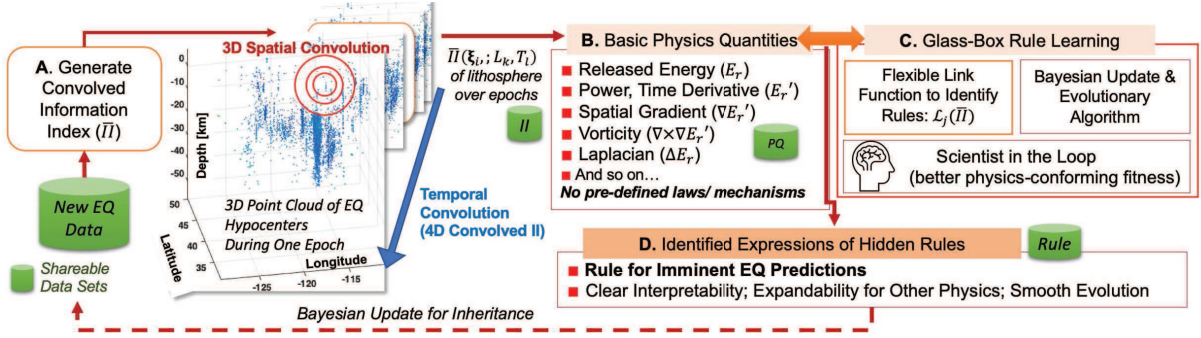}
  \caption{\textbf{Overall architecture of the glass-box physics rule learner for identifying hidden rules of imminent earthquake predictions. (A)} Externalized information convolution to generate spatio-temporal 4D convolved information index. \textbf{(B)} Data-driven basic physics (pseudo) quantities, not driven by pre-defined earthquake-related mechanisms. \textbf{(C)} Rule learning core using flexible link functions (LFs), a combination of Bayesian update and evolutionary algorithm, and scientist-in-the-loop for infusing scientists' knowledge into fitness (error) measures. \textbf{(D)} Remember the best-so-far expressions of identified rules. Shareable data sets are for other ML methods applications.}
  \label{fig:FlowChart}
\end{figure}

\begin{figure}
  \centering
  \includegraphics[width=0.90\textwidth]{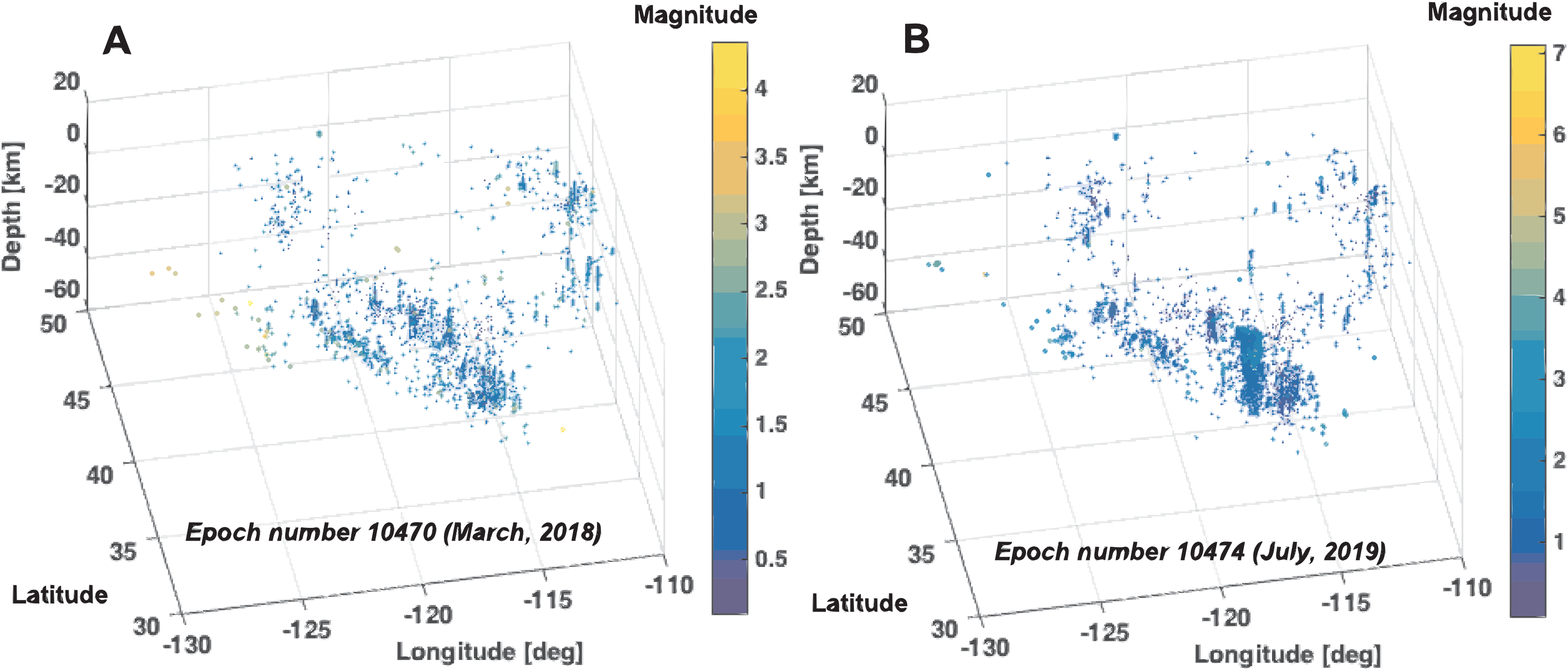}
  \includegraphics[width=0.90\textwidth]{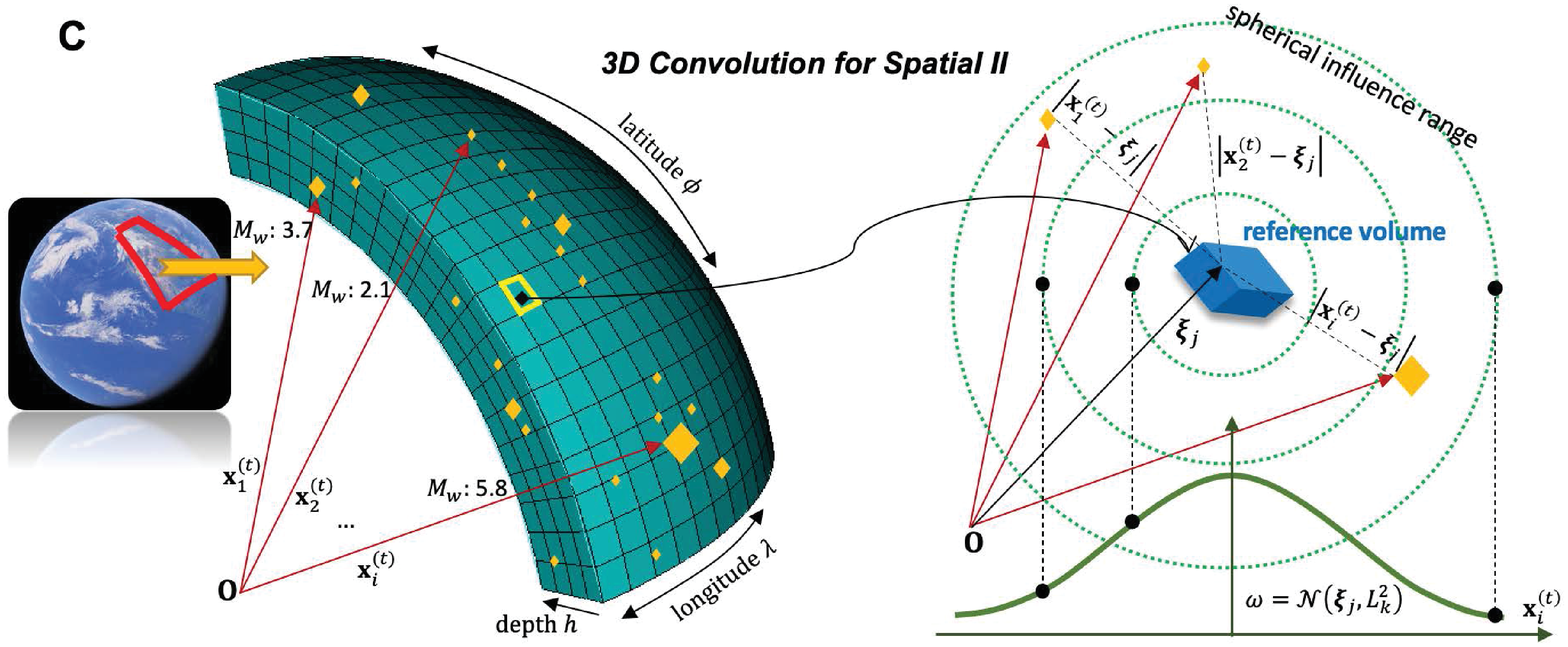}
  \includegraphics[width=0.90\textwidth]{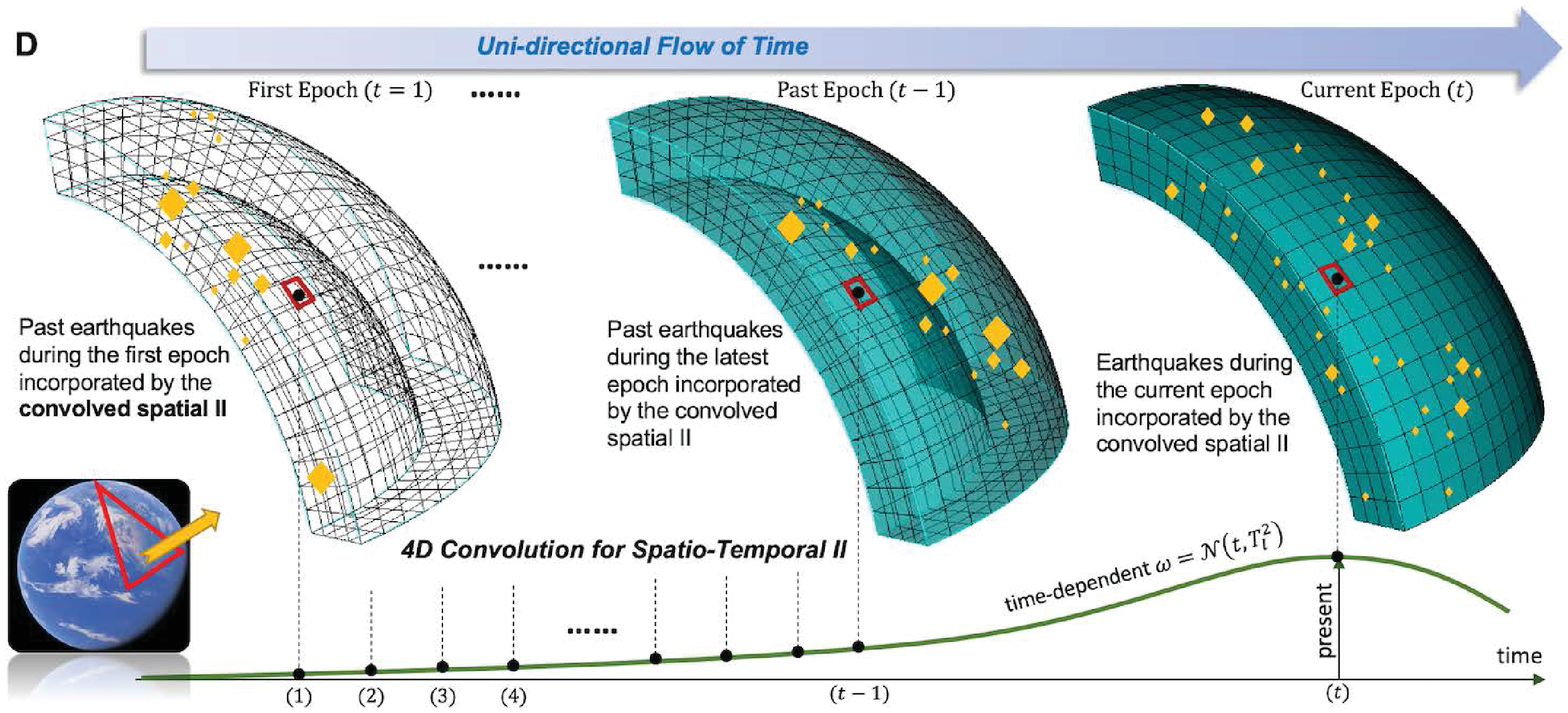}
  \caption{\textbf{Illustration of definitions of the convolved II.} \textbf{(A-B)} 3D point cloud of the recorded magnitudes of two example epochs (months) of the West region in the U.S. \textbf{(C)} Definition of the convolved spatial (3D) II. One reference volume accumulates the impacts of earthquakes during one epoch with the 3D Gaussian weights. \textbf{(D)} Convolved spatio-temporal (4D) II. Using the half Gaussian weight, all the past earthquakes are incorporated with time-decaying impacts.}
  \label{fig:Convolved_II_Definition}
\end{figure}

\begin{figure}
  \centering
  \includegraphics[width=0.9\textwidth]{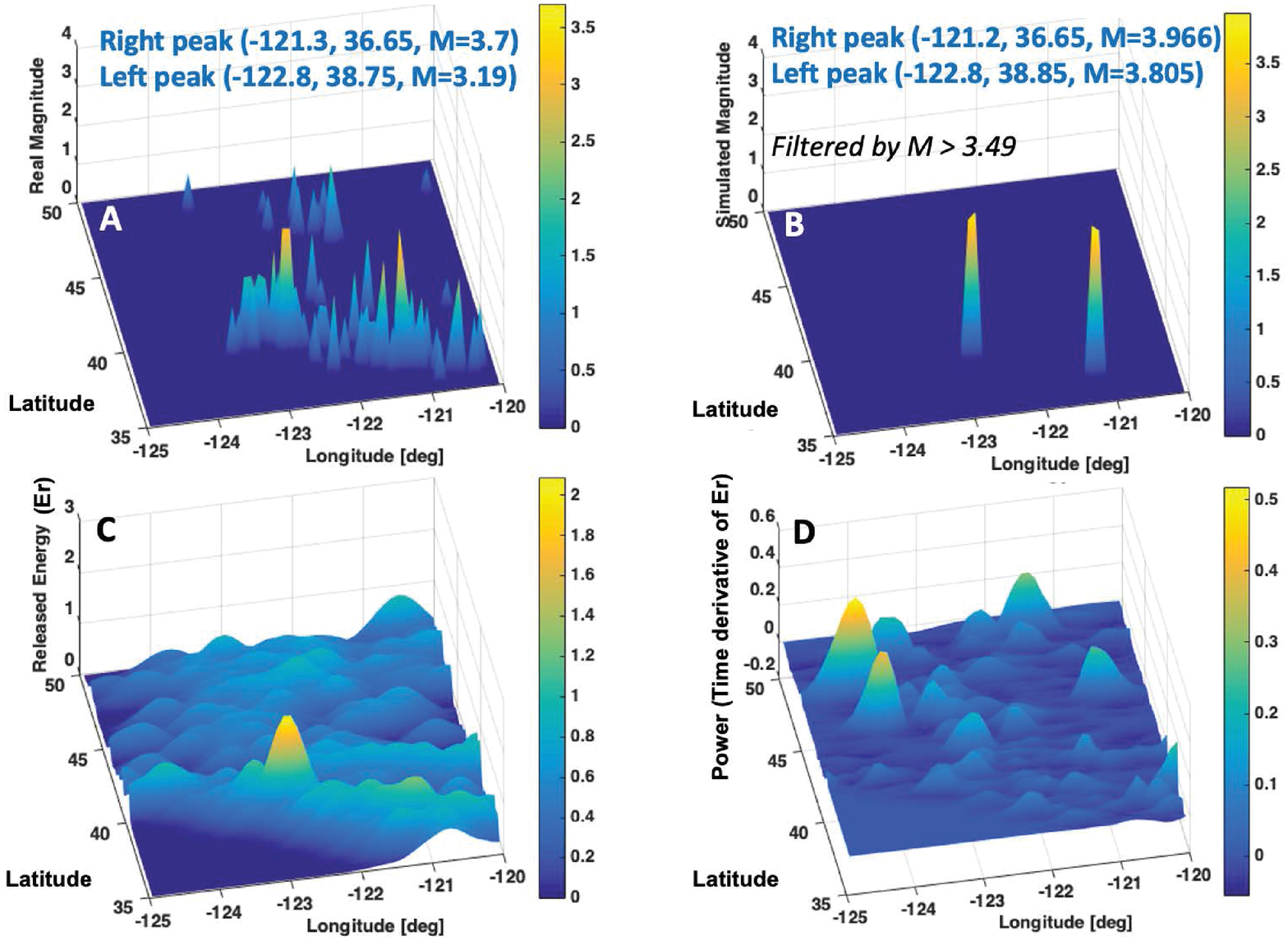}
  \includegraphics[width=0.75\textwidth]{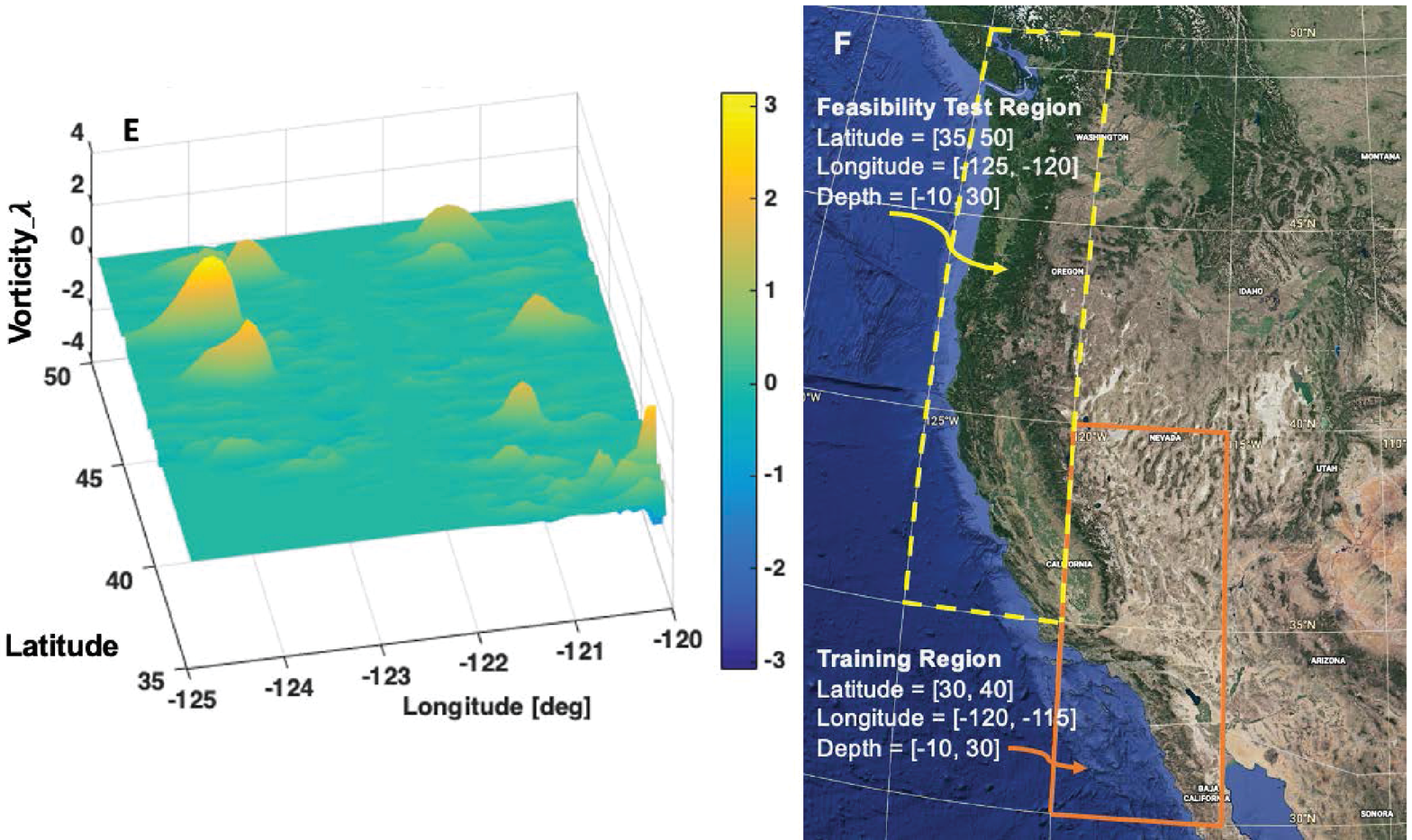}
  \caption{\textbf{Feasibility test results at depth 2.5 km with the best-so-far identified rules using the past 10 years data (epochs from 10355 through 10473) for predicting epoch 10474 (July, 2019) events.} \textbf{(A)} Observed real earthquake magnitudes at depth 2.5 km. \textbf{(B)} Predicted magnitudes using the identified rules. Only $M>3.49$ is shown since the training set the $M_{thr}=3.49$. \textbf{(C)} The released energy distribution by Eq. (\ref{eq:multiple_link_linear}). (\textbf{D}) Power distribution calculated by ($\mathcal{L}_{P}$) in Eq. (\ref{eq:Magnitude_prediction_CRS_Released_E_Power_Vorticity}). (\textbf{E}) Pseudo vorticity ($\omega_\lambda$) distribution calculated by ($\mathcal{L}_{\omega}$) in Eq. (\ref{eq:Magnitude_prediction_CRS_Released_E_Power_Vorticity}). (\textbf{F}) {The West region of the U.S. used for training and feasibility test.} (map from \it{Google Earth}).}
  \label{fig:Prediction_Test1_2_5km}
\end{figure}

\begin{figure}
  \centering
  \includegraphics[width=1.0\textwidth]{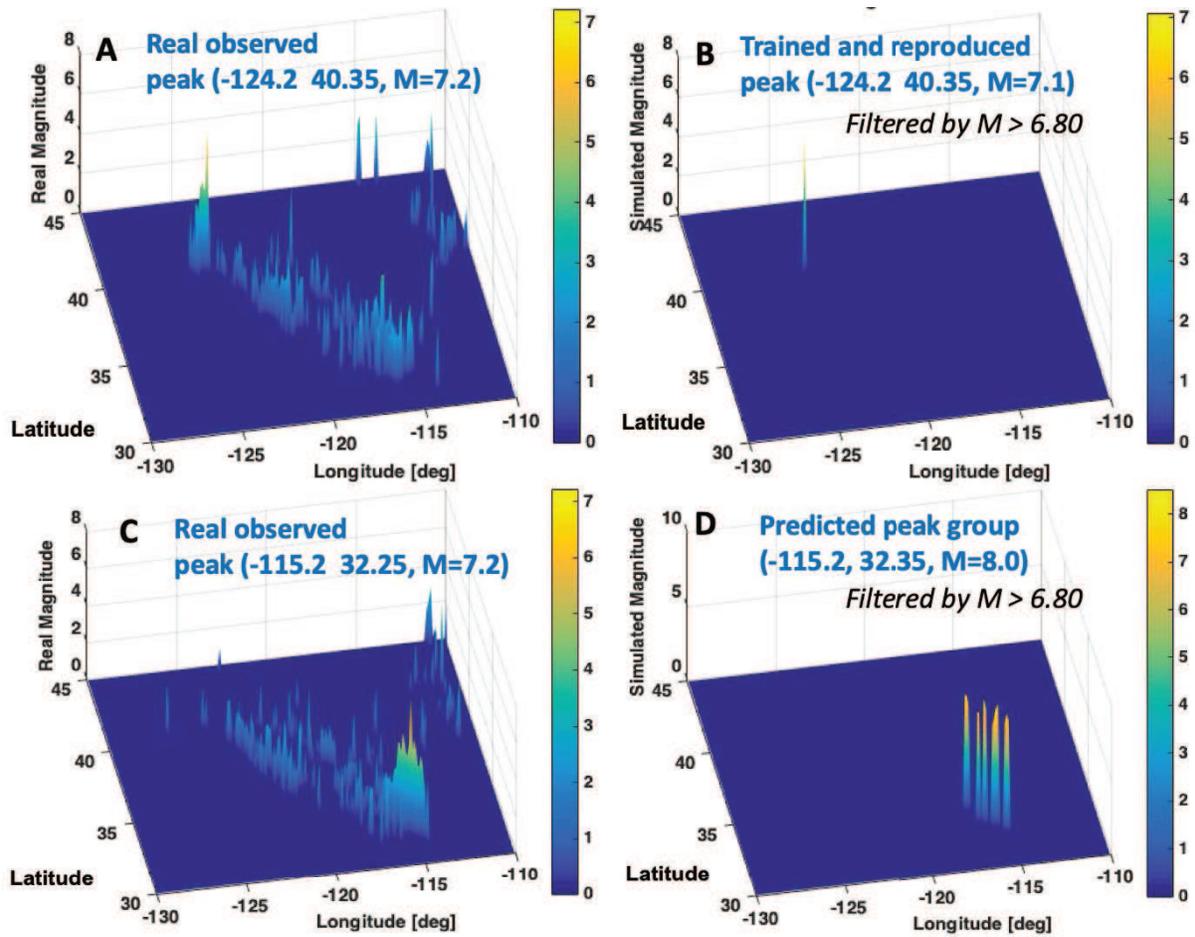}
  \caption{\textbf{Large magnitude event prediction test.} \textbf{(A)} Observed real earthquake events on April, 1992  (epoch 10147). \textbf{(B)} Trained and reproduced magnitudes using GPRL. Only $M>6.8$ is shown since this large event  application's training set the $M_{thr}=6.80$. \textbf{(C)} Observed real earthquake events on April, 2010  (epoch 10363). \textbf{(D)} Large peak group predicted by GPRL-driven rules.}
  \label{fig:Test97_PT15}
\end{figure}

\textbf{Overall architecture of the glass-box physical rule learner. }
The overall architecture of GPRL framework developed for this study is summarized in Fig. \ref{fig:FlowChart}. The central notion in Fig. \ref{fig:FlowChart} is in alignment with the author's recent applications of GPRL to the nano-scale unknown phenomena \cite{Cho:2020}. Compared to deep learning, one of the central novelties of GPRL is to ``externalize" multi-layered convolutions by conducting multiple convolutions at the information level in Fig. \ref{fig:FlowChart}A), not in the hidden layers or neurons. The starting point is raw data sets of earthquake hypocenters, of which spatial information is integrated via three-dimensional (3D) spatial convolution with multiple influence ranges (i.e. $L_{k}, k=1,..., n_l$). Then, the 3D convolved IIs are further integrated via temporal convolutions with multiple temporal influence ranges ($T_{l}, l=1,..., n_T$), thereby generating 4D spatio-temporal convolved IIs. Multiple convolved IIs and their interactions may be regarded as the counterparts to the deep learning's multi-layered convolutions. Then, scientists' basic knowledge is infused into the diverse IIs to quantify the generic terms of energy, power, gradients, or vorticity (Fig. \ref{fig:FlowChart}B). All these physics quantities are ``pseudo'' quantities since they are not from the first principle or direct physics theory. Still, they convey physical meanings. These basic quantities are derived from data via LFs and no other earthquake (EQ)-related mechanisms.  Thus, this approach pursues completely data-driven learning. Naturally, the inclusion of other physics concepts (e.g., heat, temperature, fluid) is straightforward as long as they are derived from observed data. Flexible and expressive LFs (Fig. \ref{fig:FlowChart}C) identify mathematical expressions between IIs and the basic physics quantities in Fig. \ref{fig:FlowChart}B. The revealed expressions will be about the imminent earthquake prediction as well as about physics quantities (Fig. \ref{fig:FlowChart}D). The best-so-far expressions of the identified rules are remembered and reused as a prior best generation in Fig. \ref{fig:FlowChart}D. Importantly, all the identified rules will hold clear interpretability, expandability for other physics quantities, and capability of smooth evolution. All the generated data sets (marked by the green cylinder in Fig. \ref{fig:FlowChart}) are made publicly available upon request to the author to spark independent investigations with other ML methods and catalyze innovative explorations of broad scientists.   
\textbf{Generation of convolved information index. }
The observed earthquake hypocenter data sets adopted herein \cite{USGS_Catalog} are processed into a text-based matrix form of $\{\lambda, \phi, -h, M \}_{i}^{(t)}$, $i=1,..., n^{(t)}$ and  $t=1,...,n_{ep}$ where $n^{(t)}$ means the number of total hypocenters recorded during one epoch (one month) in $[(t-1), t]$ and $n_{ep}$ means the number of total epochs (Table \ref{table:Epoch_list} summarizes the processed data from 1980 through 2019). Since this study pursues meaningful conclusions for the society, one epoch is assumed to be one month, which may be adjustable for a specific scientific reason. 
The coordinates $\{\lambda,\phi\}$ in [deg] stand for the longitude and latitude, respectively. The ground-normal $h$ is in [km], being positive above the ground datum. The magnitude $M \in [0,10)$ means the observed moment magnitude. To facilitate the spatial convolution, the geodetic coordinates $\{\lambda, \phi, -h, M \}_{i}^{(t)}$ are transformed into the earth-centered rectilinear coordinate $\{x, y, z, M \}_{i}^{(t)}$ (see \cite{Supplementary}).
A point-wise information index (II) is denoted as ``local'' II, $II_{local}\in\mathbb{R}[0, 1]$ and calculated as ${II_{local}}^{(t)}(\textbf{x}_i^{(t)}) = M_{i}^{(t)}/10$
where $\textbf{x}_i^{(t)} = (x, y, z)_i^{(t)}$.
Clearly, the local II maps real earthquake magnitudes to the range of [0,1). Fig. \ref{fig:Local_II_from10465} shows the calculated point-wise information index during the periods between epoch 10465 and epoch 10476 (i.e. from October 2018 to September 2019; \cite{USGS_Catalog}). For comparison, the raw recorded magnitudes of relatively quiet epoch (10470) and active epoch (10474) are compared in Fig. \ref{fig:Convolved_II_Definition}A-B.
Fault zones are inherently multiscale \cite{Ross:2020} with a core being surrounded by the damaged zone of which macro-fractures decay with distance from the core \cite{Mitchell:2009}. Thus, an individual earthquake's impact may not be described by a point-wise index, rather requiring a comprehensive means to capture a spatial impact on the surrounding. Complex spatial influences of many earthquakes may be integrated and accounted for by the spatial convolution presented herein. 
One of the key enablers of deep learning is the convolution process that allows information integration. If convolution is done over a spatial domain, ML can better understand the interaction of spatially distributed information and hidden patterns while applied to the temporal domain, the interactions between past and present information can be elucidated. Inheriting the philosophy of the deep learning's convolution, GPRL framework seeks to spatially integrate the local II over the 3D point cloud, i.e. myriad earthquake events in the lithosphere. The key difference from the deep learning is that this study ``externalizes" the multi-layered convolutions by conducting multiple convolutions at the information level, not in the opaque deep network layers. Rather than a uniform integration, we adopt a weighted integration using Gaussian weight function (denoted \textbf{$\omega$}) to realize the proximity-proportionate importance of information. This process generates the ``convolved spatial II" denoted as $\overline{II}_S^{(t)}$. Fig. \ref{fig:Convolved_II_Definition}C illustrates the derivation of the convolved spatial II. The physical meaning of the convolved spatial II is that $\overline{II}_S^{(t)}(\bm{\xi}_{j}; L_k)$ quantifies how much the $j$th reference volume experiences earthquakes during one epoch $(t)$ while the closer events the higher impact on the volume. The ``reference volume'' is defined as a discretized volume in the lithosphere with fixed spatial coordinate which is needed for spatial and temporal convolution (see details in \cite{Supplementary}). This study's reference volume has dimensions of (0.1 deg, 0.1 deg, 5 km) due to the limit of computational resources. If the earthquakes during the epoch took place nearby (i.e. within or close to the $L_k$) the failure directly affects the $j$th reference volume whereas earthquakes occurred at distance (i.e. much larger than $L_k$), the reduced impact is recorded in the $j$th reference volume via the  $\overline{II}_S^{(t)}(\bm{\xi}_{j}; L_k)$.
This is a time-dependent quantity and thus defined at an epoch $(t)$ and calculated as  
\begin{equation}\label{eq:nonlocal}
\overline{II}_S^{(t)}(\bm{\xi}_{j}; L_k) = \int_{\text{V}} { \omega(\bm{\xi}_j, \textbf{x}_{i}^{(t)}; L_k) II_{local}^{(t)}(\textbf{x}_{i}^{(t)}) d\textbf{x}}
\end{equation}
where $\omega(\bm{\xi}_j, \textbf{x}_{i}^{(t)}; L_k) = (L_k(2\pi)^{1/2})^{-N}\exp\left(-\frac{|\textbf{x}_{i}^{(t)} - \bm{\xi}_j|^2}{2L_k^2}\right) = \mathcal{N}(\textbf{x}_{(i)}^{(t)}, L_k^2)$; $\bm{\xi}_{j}$ stands for the position vector of the center of $j$th reference volume and $\forall \textbf{x}_{i}^{(t)} \in \text{V}$, and V means the entire lithosphere domain under consideration. The detailed scheme for discretization of the integration Eq. (\ref{eq:nonlocal}) is presented in \cite{Supplementary}. 
In Eq. (\ref{eq:nonlocal}), $L_k \in \mathbb{R}^{+}$, $k = 1,..., n_L$ stands for the radius of influence range. With a larger value of $L_k$, the earthquake events across a broad space can be incorporated at the expense of over-smoothing effect; with a smaller $L_k$, higher priority on the adjacent earthquakes to the current reference volume at the expense of local spikes or over-fitting effect. For the weighting function, there is no restriction to the use of other weightings. The dimension parameter $N=3$ is used for the spatial convolution over 3D point cloud whereas $N=1$ is used for the temporal convolution over time which shall be explained later.
Fig. \ref{fig:Spatial_II_z12_5km} shows three cases of the convolved spatial II with different influence ranges. All three cases used the procedures given in Eq. \ref{eq:nonlocal}. Still, with a larger $L>50$ km such as Fig. \ref{fig:Spatial_II_z12_5km}C, the over-smoothing effect is notable. In heterogeneous materials or composite structures, this spatial convolved II may help ML understand internal complexity as scientists do \cite{Cho:2019}.
\newline
Earthquake is not a one-time event, but a reference volume in the lithosphere experiences incessantly many events over time. By extending convolution to the time domain, we can incorporate such transient information about how one reference volume has been being affected by past earthquakes. Performing convolution over time creates ``convolved spatio-temporal II" (denoted as $\overline{II}_{ST}^{(t)}$). This convolved spatio-temporal II accounts for all the past earthquakes up to the present epoch $t$. Fig. \ref{fig:Convolved_II_Definition}D illustrates the calculation procedure of the spatio-temporal II. Since it embraces information over space and time, its another name would be 4D II. The one-dimensional ($N=1$) Gaussian weighting is used, being centered at the present time $t$. Being not certain about the optimal temporal influence ranges, here we allow in total $n_T$ temporal influence ranges, denoted by $T_l, l=1,...,n_T$. For a temporal influence range $T_l$, we have
\begin{equation}\label{eq:nonlocal-spatio-temporal}
\overline{II}_{ST}^{(t)}(\bm{\xi}_{j}; L_k, T_l) = \int { \omega(\tau; T_l) \overline{II}_S^{(t_{past})}(\bm{\xi}_{j}; L_k) dt_{past}}
\end{equation}
where $\omega(\tau; T_l) = (T_l(2\pi)^{1/2})^{-1}\exp\left(-\frac{\tau^2}{2T_l^2}\right) = \mathcal{N}(t, T_l^2)$; $\tau = |t-t_{past}|, t \geq t_{past}$, meaning the time gap between the current and the past time, all given in [epoch]. This convolved spatio-temporal II is calculated at the $j$th reference volume center, $\bm{\xi}_{j}$. 
As in the spatial convolved II, this spatio-temporal II is generated by the discretization scheme described in \cite{Supplementary}.
Importantly, in the discretization scheme, the incremental time-lapse $\Delta t_i$ over one epoch, which is assumed to be constant $\Delta t_i = 1$ epoch (i.e. one month). With a larger value of $T_l$, the earthquake events across a long past period can be incorporated at the expense of over-smoothing effect; with a smaller $T_l$, a higher priority on the most recent earthquakes to the present time at the expense of local spikes or over-fitting effect. Choosing values and total counts of $T_l$ is subject to learning and prediction accuracy. In concept, this inclusion of temporal effects with time-dependent decaying influence shares the similar notions of the well-known long short-term memory (LSTM) and its variants \cite{Hochreiter:1997, Gers:2000} that uses selective remembering and forgetting in the time axis.
\newline
Fig. \ref{fig:Temporal_Convolution_Illustration2} explains the training with an epoch time frame. Departing from the first epoch, all the spatio-temporal convolved IIs within the epoch frame will be used for training and then the identified rule is used to predict the earthquakes of the last epoch of the frame \cite{Supplementary}.
As expected, the convolved spatio-temporal II appears to successfully quantify earthquake events (Fig. \ref{fig:II_ST_12epochs_from10462_or_10465}) and effectively distinguish the low and high seismic activities. The convolved spatio-temporal II may span substantially small values which may hamper the searching power of the adopted link function (LF). For instance, the cubic regression spline (CRS)-based LF takes a covariate ranging between [0,1], and thus if the covariate is very small the subsequent learning may not be efficient. Thus, it is helpful to rescale the convolved spatio-temporal II to [0,1]. Details about the recommended normalization scheme and a succinct proof of the upper bound are presented in \cite{Supplementary}.
\newline
\textbf{Flexible and transparent link functions. }
Placing top priority on the interpretability, this study proposes to adopt an expressive link function (LF) using transparent, flexible basis that can describe a mathematical expression between the convolved spatio-temporal II, $\overline{II}_{ST}$ and the hidden physical rules. LF is denoted as $\mathcal{L}(\overline{II}_{ST}; \bm{\uptheta})$ where $\bm{\uptheta}$ is a set of free parameters prescribing the LF. This study used an evolutionary algorithm coupled with the Bayesian update scheme to enable LF to continue to learn, train, and evolve. There is little restriction of choice of other forms of LFs. For balancing the efficiency and interpretability, one may choose the cubic regression spline (CRS)-based LF with high flexibility \cite{Wood:2006, Gu:2013} or two-parameter based exponential LF with its simplicity. First, the CRS-based LF has a general form as  $\mathcal{L}^{(k,l)}(\overline{II}_{ST}^{(t)}(\bm{\xi}_{j}; L_k, T_l) ; ~\bm{\uptheta}^{(k,l)}) = \sum_{i=1}^p{a_{i}^{(k,l)} b_{i}^{(k,l)}(\overline{II}_{ST}^{(t)}(\bm{\xi}_{j}; L_k, T_l))}$ where $\bm{\uptheta}^{(k,l)} = \{\textbf{a}, \textbf{x}^*\}^{(k,l)}$ with $\textbf{a}^{(k,l)} = \{a_1,...,a_p\}^{(k,l)}$, the knots $\textbf{x}^{*(k,l)} = \{x_1^*,...,x_{(p-2)}^*\}^{(k,l)}$, and the cubic spline basis $b_{i}^{(k,l)}$ given in Eq. (\ref{eq:cubic_spline}) in \cite{Supplementary}.
Next, the two-parameter exponential LF has a simpler form as $\mathcal{L}^{(k,l)}(\overline{II}_{ST}^{(t)}(\bm{\xi}_{j}; L_k, T_l) ; ~\bm{\uptheta}^{(k,l)}) = \text{exp}\left(a^{(k,l)}\overline{II}_{ST}^{(t)}(\bm{\xi}_{j}; L_k, T_l)^{b^{(k,l)}}\right) - 1$ where $\bm{\uptheta}^{(k,l)} = \{a^{(k,l)}, b^{(k,l)}; k=1,...,n_L, l=1,...,n_T\}$, and ``-1" is to make the minimum of the LF near zero. It should be noted that the exponential LF is always non-zero, positive, and monotonically increasing while preserving the concave or convex shape (see Fig. \ref{fig:exponential_LF} in \cite{Supplementary}).  
\newline
\textbf{Hidden expression of the released energy. } 
By leveraging the flexible and expressive LFs, the observed earthquake data help identify a hidden expression about the released energy in the lithosphere. Earthquakes leave behind a footprint on energy. For instance, earthquakes result in macroscopic and microscopic damages and fractures on the surrounding solid domain \cite{Ross:2020, Mitchell:2009} as well as the faults. The released energy (denoted as $E_r^{(t)}(\bm{\xi}_{j}) \in \mathbb{R}^{+}$) of the $j$th reference volume at current time $t$ may be represented in terms of the convolved spatio-temporal IIs. It should be noted that this paper does not adopt the well-proven magnitude-energy power law since the present goal is to establish a purely data-driven hidden rule learning. Owing to the accumulated influences of adjacent earthquakes over time, it is plausible to consider that the released energy at a reference volume is increasing. Thus, the simple exponential LF is preferred for the hidden relation between the released energy and the convolved spatio-temporal II. The true physical rule of the released energy, if exists, is hard to formulate, and thus this approach seeks to borrow the learning power of GPRL. Amongst many possible combination operations (e.g., $+$ or $\times$), the additive operation is found to be favorable. 
The identified best-so-far expression of the released energy is given by 
\begin{equation}\label{eq:multiple_link_linear}
E_r^{*(t)}(\bm{\xi}_{j}) = \text{max}\left[\sum_{k=1}^{n_L=2} \sum_{l=1}^{n_T=2} \mathcal{L}^{(k,l)}(\overline{II}_{ST}^{(t)}(\bm{\xi}_{j}; L_k, T_l) ; ~\bm{\uptheta}^{(k,l)}), 0.0\right]
\end{equation}
where the best-so-far free parameters $\bm{\uptheta}^{(k,l)}$ are summarized in Table \ref{table:final_free_parameters_LFs}. Example plots using the exponential LFs with additive combination are shown in Fig. \ref{fig:Released_E_Exponential}. And the best combination of the spatial and temporal influence ranges are identified as $L_k=(10, 25)$ [km] and $T_l=(3, 6)$ [epoch = month] by comparative investigations. This combination of short- and long-range influence ranges appears to outperform the other rules with a single $L$ or $T$ or many $L$'s and $T$'s. As shown in Fig. \ref{fig:LF_Er_S_CA_Test48}, relative contribution of different influence ranges appears complicated but interpretable. In the higher II ranges ($\overline{II}_{ST} >$ 0.8), the spatio-temporal II with $(L_1, T_2)$ = (10 km, 6 epochs) and $(L_2, T_1)$ = (25 km, 3 epochs) to the released energy are significant (see Figs. \ref{fig:LF_Er_S_CA_Test48}B-C). In contrast, the contribution of II with $(L_1, T_1)$ = (10 km, 3 epochs) are uniform regardless of $\overline{II}_{ST}$ and thus important in the low and mid ranges of II ($\overline{II}_{ST} <$ 0.8; Fig. \ref{fig:LF_Er_S_CA_Test48}A). Although this identified rule of the released energy may not be close to the ``exact'' one, the clear interpretability of the identified rule is still meaningful, conveying physically-sound implications. For instance, Fig. \ref{fig:LF_Er_S_CA_Test48}A implies that nearly all the earthquakes in close distance and recent time retain their influence. Contrarily, Figs. \ref{fig:LF_Er_S_CA_Test48}B-C imply that only larger earthquakes ($\overline{II}_{ST}>$0.8) retain influence because they are far away or old enough to allow post-earthquake curing.    
\newline
\textbf{Pseudo power and pseudo vorticity of the released energy. }
Other important physics quantities would be the spatial gradients of the released energy over the lithosphere and ``power.'' The time derivative of energy is physically related to the power. The calculation procedure of the time derivatives of the energy-related terms is presented in \cite{Supplementary}. Figs. \ref{fig:PT1_z2_5km_Grad} and \ref{fig:PT1_z12_5km_Grad} present example plots of the spatial gradients and time derivative of the released energy at depth 2.5 km and 12.5 km, respectively. These plots are with respect to the earth-centered coordinate system before transformation to geodetic coordinates. 
The spatial gradient with respect to the geocentric coordinate system may convey weak physical and geometrical information in view of the curved structure of the earth lithosphere. Thus, it is meaningful to transform the geocetric gradient (denoted as $\nabla{E}_{r}^{(t)}$) to the geodetic gradient (denoted as $\nabla_{g} {E}_{r}^{(t)}$), i.e., the spatial gradient with respect to the geodetic coordinate system (${\lambda, \phi, h}$). This can be done by Jacobian \textbf{J} (details are in \cite{Supplementary}): $\nabla_{g} E_r^{(t)}(\bm\xi_j) = \textbf{J}\nabla E_r^{(t)}(\bm\xi_j)$. 
Fig. \ref{fig:Spatial_Grad_Released_E} shows example plots of the gradient field vector at depth $z$ = 12.5 km with respect to the geodetic coordinate system.
By observing the transient change of the spatial gradient of the released energy, this study derives the pseudo ``vorticity'' $\bm{\omega} = (\omega_\lambda, \omega_\phi, \omega_h)$ as  
\begin{equation}\label{eq:Vorticity}
\bm{\omega} := \nabla_g \times \left( \nabla_g{ \frac{\partial E_r^{(t)}(\bm{\xi}_{j})}{\partial{t}}} \right)
\end{equation}
\begin{equation}\label{eq:Vorticity_detail}
= \left( \frac{\partial}{\partial\phi}\frac{\partial{E_r^{'}}}{\partial{h}} - \frac{\partial}{\partial{h}}\frac{\partial{E_r^{'}}}{\partial\phi}, \: \frac{\partial}{\partial{h}}\frac{\partial{E_r^{'}}}{\partial \lambda} - \frac{\partial}{\partial \lambda}\frac{\partial{E_r^{'}}}{\partial{h}}, \:  \frac{\partial}{\partial\lambda}\frac{\partial{E_r^{'}}}{\partial\phi} - \frac{\partial}{\partial\phi}\frac{\partial{E_r^{'}}}{\partial\lambda} \right)
\end{equation}
In Eq. (\ref{eq:Vorticity}), $\nabla_g=(\partial/\partial\lambda; \partial/\partial\phi, \partial/\partial{h})$; "$\times$" is the curl operator; $E_r'=\frac{\partial E_r^{(t)}(\bm{\xi}_{j})}{\partial{t}}$. 
Fig. \ref{fig:PT1_z2_5km_Vorticity} presents example plots of the calculated vorticity vector.
The vorticity of the released energy flow is considered as another physics quantity since the vorticity may hint at the temporal rotation of the strain energy field which may play an important role in rupture initiation. There is no direct definition of the velocity field needed for vorticity calculation, and thus the spatial gradient of the time derivative of the released energy ($\nabla_g{\frac{\partial E_r^{(t)}(\bm{\xi}_{j})}{\partial{t}}}$) is regarded as a ``pseudo velocity'' in Eq. (\ref{eq:Vorticity}). Physically, this pseudo velocity field may describe the spatial distribution of how the released energy is changing over time. Although the time increment is large (here, one month) compared to mathematical derivative, the slow motion of the earth plate (e.g., 8-10 cm/year \cite{Simon:2011}) may justify the use of such a large time interval for the pseudo velocity. 
%
\newline
\textbf{The best-so-far identified rule of magnitude prediction. } 
Without any prior knowledge of existing magnitude prediction models (e.g., \cite{Rundle:2003, Sotolongo-Costa:2004, Keilis-Borok:2003, Tiampo:2002}), this study directly seeks to find a hidden rule of magnitude predictions. Aiming at a purely data-driven pathway, this study explores basic physics quantities that can be derived from the observed data.
Training and searching for the best-performing rules are conducted by GPRL framework on the West-South region of the U.S. (solid box in Fig. \ref{fig:Prediction_Test1_2_5km}F). Using the identified rule, the separate feasibility test was conducted on the West-North region of the U.S. (dashed box in Fig. \ref{fig:Prediction_Test1_2_5km}F). To ensure independent feasibility test, no data of the training region are used for the feasibility test and vice versa. In pursuit of the best rules, many possible candidates of physics quantities are explored: the released energy $E_r$ and many forms of physical variants of $E_r$ including the three components of the spatial gradient vector $\nabla_{g} {E}_{r}^{(t)}$, the local maximums and minimums of $\nabla_{g} {E}_{r}^{(t)}$, the time derivative $\partial\nabla_{g} {E}_{r}^{(t)}/\partial{t}$ meaning the power, and the pseudo vorticity (Eq. \ref{eq:Vorticity}).
From the comparative investigations, the best-so-far rule of the magnitude prediction identified by GPRL framework suggests holding three physics quantities: (1) the released energy (the corresponding best-so-far CRS LF is denoted by $\mathcal{L}_{E}$), (2) the power, i.e., the time derivative of the released energy ($\mathcal{L}_{P}$), and (3) the pseudo vorticity of the released energy flow ($\mathcal{L}_{\omega}$). 
The best-so-far rule of magnitude prediction is identified as the multiplicative combination of these CRS LFs of three physics quantities as    
\begin{equation}\label{eq:Magnitude_prediction_CRS_Released_E_Power_Vorticity}
M_{pred}^{(t+1)}(\bm{\xi}_j) := \mathcal{L}_{E}(E_r^{*(t)};~ \bm\uptheta_E) \mathcal{L}_{P}\left( \text{Sg}\left( e^2 \frac{\partial E_r^{*(t)}}{\partial{t}} \right) ;~ \bm\uptheta_P\right) \mathcal{L}_{\omega}\left( \text{Sg}\left( e^2\omega_{\lambda} \right) ;~ \bm\uptheta_\omega\right)
\end{equation}
where $E_r^{*(t)}$ is the best-so-far rule-driven released energy at epoch $t$ and at the reference volume $\bm{\xi}_j$. The best-so-far free parameters $\bm{\uptheta}_{E}$, $\bm{\uptheta}_{P}$, and $\bm{\uptheta}_{\omega}$ are summarized in Table \ref{table:final_free_parameters_LFs}. $\text{Sg}(.)$ stands for a typical sigmoid function, $\text{Sg}(x) = 1/(1+e^{-x})$, for brevity. The power term's LF $\mathcal{L}_{P}$ uses the sigmoid function to transform $\partial E_r^{(t)}(\bm{\xi}_{j})/\partial{t} \in \mathbb{R}[-\infty, \infty]$ to $\mathbb{R}(0,1)$ which is compatible with the input range of CRS bases. A slightly modified sigmoid with a scaling-up factor $e^2$ is used since it appears to outperform against a typical sigmoid case. This scheme applies to the pseudo vorticity's LF $\mathcal{L}_{\omega}$  since $ \omega_h \in \mathbb{R}[-\infty, \infty]$. 
Amongst many candidates for $\mathcal{L}_{\omega}$, e.g., $\omega_{\lambda}, \omega_{\phi}, \omega_{h}, \text{or} \sqrt{\omega_{\lambda}^2 + \omega_{\phi}^2}$, comparative investigations suggest that $\omega_{\lambda}$ appears to give the most plausible performance, as finally included in Eq. (\ref{eq:Magnitude_prediction_CRS_Released_E_Power_Vorticity}). Physically, $\omega_{\lambda}$ may describe the slow rotational motion of the energy flow about the longitudinal axis. This study's training data are from the Southern U.S. region of which plate motions and the known major faults are roughly parallel or normal to the longitudinal axis. This coincidence may underpin the relatively important role of $\omega_{\lambda}$ in the identified rule of magnitude prediction.    
The feasibility test results at depth 2.5 km are shown in Fig. \ref{fig:Prediction_Test1_2_5km} which are generated by using the best-so-far identified rules by GPRL that are based on the spatial and temporal influence ranges, i.e. $L_j \in  \{10, 25\}$ [km] and  $T_k \in  \{3, 6\}$ epochs (months). The underlying spatio-temporal 4D convolved IIs are presented in Fig. \ref{fig:PT1_z2_5km_II_ST}. The best-so-far prediction rule appears to reproduce two peaks near the real two peaks' locations (Fig. \ref{fig:Prediction_Test1_2_5km}A-B). It should be noted that the prediction rule training and error function are all based on the minimum magnitude threshold $M_{thr}=3.49$ (Fig. \ref{fig:Prediction_Test1_2_5km}B). Interestingly, from the naked eyes, the spatial irregularities in the 4D IIs are not apparent. But their spatial and transient fluctuations gradually emerge and become noticeable as the 4D IIs are transformed into other physics quantities of the released energy, its spatial gradients (Figs. \ref{fig:PT1_z2_5km_Grad}A-C), the time derivative of the spatial gradients of the released energy (power-like quantity; Figs. \ref{fig:PT1_z2_5km_Grad}D-F), and the pseudo vorticity (Figs. \ref{fig:PT1_z2_5km_Vorticity}A-C).
\begin{figure}
  \centering
  \includegraphics[width=1.0\textwidth]{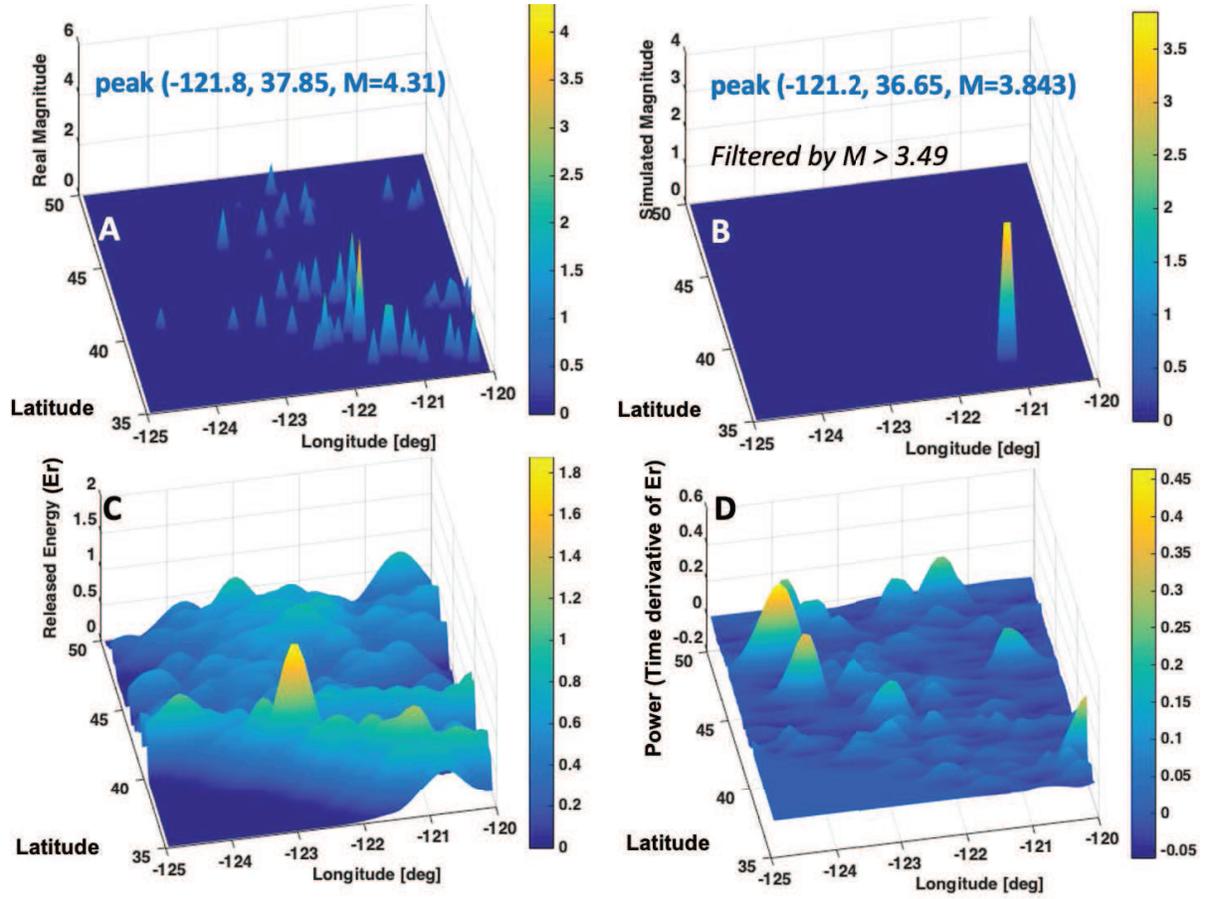}
  \caption{\textbf{Feasibility test results at depth 12.5 km with the best-so-far identified rules using the past 10 years data (epochs from 10355 through 10473) for predicting epoch 10474 (July, 2019) events.} \textbf{(A)} Observed real earthquake magnitudes at depth 12.5 km. \textbf{(B)} Predicted magnitudes using the identified rules. Only $M>3.49$ is shown since the training set the $M_{thr}=3.49$. \textbf{(C)} The released energy distribution by Eq. (\ref{eq:multiple_link_linear}). (\textbf{D}) Power distribution calculated by ($\mathcal{L}_{P}$) in Eq. (\ref{eq:Magnitude_prediction_CRS_Released_E_Power_Vorticity}). Pseudo vorticity ($\omega_\lambda$) distribution is shown in Fig. \ref{fig:PT1_z12_5km_Vorticity}. }
  \label{fig:Prediction_Test1_12_5km}
\end{figure}
Another feasibility test result at a different depth 12.5 km where the overall peak ($M=4.31$) took place is shown in (Fig. \ref{fig:Prediction_Test1_12_5km}). The underlying spatio-temporal 4D convolved IIs are presented in Fig. \ref{fig:PT1_z12_5km_II_ST}. As in the prediction results at depth = 2.5 km, despite the weak variations in the spatio-temporal IIs, other physics quantities appear to magnify the hidden spatio-temporal fluctuations inside the lithosphere at depth 12.5 km (see Figs. \ref{fig:PT1_z12_5km_Grad} and  \ref{fig:PT1_z12_5km_Vorticity}). Despite the error in magnitude and location, the best-so-far prediction rule appears to pinpoint the single peak near the real earthquake event (Fig. \ref{fig:Prediction_Test1_12_5km}A-B).  
\newline
\textbf{Application to Large Magnitude Event Prediction (M $>$ 7.0). } 
As an application to large earthquake prediction, the best-performance setting of GPRL of the previous sections is applied to training of large event of magnitude larger than 7.0 in the entire West region of the U.S. (i.e., longitude and latitude in (-130, -110) and (30, 45) [deg], respectively, and depth (-5, 20) [km]). GPRL is trained with past 10 years earthquake data up to March 1992 with training target (epoch 10147) magnitude 7.2 earthquake occurred on April 25, 1992 at (40.3353333 deg, -124.22867 deg, 9.856 km). From this training, the best-so-far rule identified by GPRL appears to be successful in reproducing the next-month earthquake's location and magnitude as shown in Fig. \ref{fig:Test97_PT15}A-B. Then, with the learned best-so-far rules, an independent prediction test is conducted to predict magnitude 7.2 earthquake (epoch 10363) on April 4, 2010 at (32.2861667 deg, -115.2953333 deg, 9.987 km). As shown in Fig. \ref{fig:Test97_PT15}C-D, the best-so-far identified rule of GPRL appears to successfully predict the location and magnitude of the large event on April, 2010, notably using the observed 10 years data 30 days before the event without any physics mechanisms or statistical laws. It is interesting to note that the GPRL-driven rule appears to predict localized large event (Fig. \ref{fig:Test97_PT15}B) in a relatively narrow zones and also the group of large activities (Fig. \ref{fig:Test97_PT15}D) across relatively wide zones. The associated results of released energy, pseudo power, and pseudo vorticity are presented in Fig. \ref{fig:Test97_PT15_pseudo}.      
\section*{Discussion}
This study proposed a mere data-guided pathway to the long-sought rule of imminent ``individual'' earthquake predictions.
To seek fundamentally different data-driven approach, this paper focuses on predicting individual future earthquake's location and magnitude instead of collective event counts or overall probability, and this paper intentionally uses the observed earthquake data without adopting any pre-defined statistical laws such as GR law, Omori law, fertility law, or magnitude-energy law. Without any prejudice of earthquake-related mechanisms or statistical laws, the observed data of decades-long hypocenters appear to help unravel the hidden rule of predicting imminent individual earthquake events before 30 days. Amongst many physics quantities, the best-so-far rule is identified to hold the pseudo power and pseudo vorticity (about longitudinal axis) of the released energy in the lithosphere. The identified rules appear to have a complex nonlinear relationship with the internal physics quantities, which underpins the strong learning capability of the proposed GPRL framework. Unlike the black-box machine learning, the adopted glass-box physics rule learning offers an interpretable expression of the identified rule as expected. Independent feasibility test supports a promising role of the proposed method. Still, the identified rule and GPRL framework are not the final version but an initial version, being subject to substantial improvement and evolution. Since its starting point is the observed data, the improvement of the earthquake data sets \cite{Ross:2018, Pardo:2019, Xu:2019} will positively influence the prediction accuracy. As long as the reliability and precision of the relevant data is ensured, further inclusion of more physics (e.g., thermal instability \cite{Wang:2020}, pore pressure \cite{Zhu:2020}, fluid injection \cite{Ross:2020}) into the present framework would lead to a positive improvement, which will be straightforward in view of clear interpretability and extensibility of the present framework.
With respect to computational aspects of GPRL, there is ample room for further sophistication. For instance, it would be beneficial to consider more flexible, versatile bases \cite{Wood:2006} for the link functions, an extensive library of possible mathematical expressions \cite{Champion_et_al:2019}, powerful symbolic regression methods \cite{Udrescu:2020}, or stochastic optimizer \cite{Kingma:2015}. Consistent evolution or automated optimization of many hyper-parameters of the GPRL framework may be done by inheriting the reinforcement learning paradigm \cite{Sutton:2017}.
In light of the multifaceted nature of earthquake phenomena, enabling imminent individual earthquake predictions will require comprehensible collaborations of geophysics, mechanics, computer science, data science, and so on like a recent multi-disciplinary global collaborations \cite{Mignan:2021}. The initial outcome of this study catalyzes such a broad endeavor. Notably, this paper paves an initial ground for purely data-driven prediction of individual large earthquake's location and magnitude one month before, and its potential is boundless for science and humanity.    
\section*{Methods}
\textbf{Raw earthquake data sets of training epochs. }
This study collected and processed of raw earthquake catalog data available in \cite{USGS_Catalog} between January 1980 and October 2019 as summarized in Table \ref{table:Epoch_list}. One training epoch corresponds to one month's earthquake events -- epoch 10000 stands for January 1980, epoch 10477 for October 2019. All of these earthquake data sets are made available upon request to the author. 
\newline
\textbf{Computational implementation of proposed algorithms. } The spatio-temporal convolution of earthquake catalog data to generate the convolved information index is computationally expensive. This study developed a parallelized glass-box physics rule learner framework with \textit{C++} and \textit{OpenMPI}. All other learning, evolutionary algorithm and Bayesian update scheme are implemented on the parallel program. The developed program is made available upon request to the author. Iowa State University's high-performance computing facility, \textit{Condo} cluster is used for this study. Tables \ref{table:train_setup}, \ref{table:train_setup2}, \ref{table:feasibility_test_setup}, and \ref{table:feasibility_test_setup2} present the input setup of GPRL program used for training and feasibility test.  
\newline
\textbf{Reference volumes in the Earth lithosphere domain. }
This study defines reference volumes of the given domain in the Earth lithosphere. Within a time period, The raw data of earthquake hypocenters are distributed over the lithosphere domain and thus constitute a sort of irregular 3D point cloud. To generate 4D convolved spatio-temporal information index (II) by performing spatial and temporal convolutions, it is efficient to define a fixed location in the space and time, which is the central reason for the reference volumes. Admittedly the Earth lithosphere is not a simple spherical structure, and the earthquake hypocenters are often recorded on longitude, latitude and depth, $(\lambda,\phi,h)_i^{(t)}, i=1,..., n^{(t)}$. Therefore, this study processes raw data to the earth-centered 3D coordinates, $(x,y,z)_i^{(t)}, i=1,..., n^{(t)}$.   

\begin{enumerate}
    \item {Transform Raw Hypocenter Data to Geocentric Coodinates:} At current epoch $(t)$, the first step is to read each hypocenter's raw coordinate $(\lambda,\phi,-h)_i^{(t)}$ where the longitude $x\lambda\in [-180, 180]$ is in [deg], the latitude $\phi \in [-90, 90]$ in [deg], and the depth $h$ in [km]. Here, $h$ means the ellipsoidal height along its normal, being positive outward normal to the reference ellipsoid. Note the earthquake catalog data use the reversed sign convention of $h$. Transform them to the earth-centered 3D spatial coordinate $(x, y, z)_i^{(t)}$ (i.e. geocentric rectangular coordinates) as described in \cite{Lichtenegger:2012}
    \begin{equation}\label{eq:earth-centered_x}
    x = \left(\frac{a^2}{\sqrt{a^2 \text{cos}^2 \phi + b^2 \text{sin}^2 \phi}} + h \right)\text{cos}\phi \; \text{cos}\lambda
    \end{equation}    
    \begin{equation}\label{eq:earth-centered_y}
    y = \left(\frac{a^2}{\sqrt{a^2 \text{cos}^2 \phi + b^2 \text{sin}^2 \lambda}} + h \right)\text{cos}\phi \; \text{sin}\lambda
    \end{equation}
    \begin{equation}\label{eq:earth-centered_z}
    z = \left(\frac{b^2}{\sqrt{a^2 \text{cos}^2 \phi + b^2 \text{sin}^2 \lambda}} + h \right)\text{sin}\phi
    \end{equation}
    where $a=6378.1370$ km and $b=6356.7523$ km according to the 1984 World Geodetic System (WGS 84) revision. 
    \item {{Reference Volumes:} Given the ranges of longitudes, latitudes, and depths, this study defines uniformly distributed grid system and each cell is denoted as a reference volume. The total number of reference volumes $n_{rv}$ is simply calculated by $n_{rv} = n_{\lambda} \times n_{\phi} \times n_{h}$ where $n_{\lambda} = |\lambda_{max} - \lambda_{min}|/\Delta{\lambda}$, $n_{\phi} = |\phi_{max} - \phi_{min}|/\Delta{\phi}$, and $n_{h} = |h_{max} - h_{min}|/\Delta{h}$. Here $ (\Delta{\lambda}, \Delta{\phi}, \Delta{h}) $ are user-defined increments of longitude, latitude, and depth, respectively. The index of reference volume is ordered by $\lambda, \phi,$ and $h$. Thus, the coordinates or the $j$th reference volume's center, denoted as $\bm{\xi}_j\in \mathbb{R}^3$, is represented by 
    \begin{equation}\label{eq:ref_coor}
    \bm{\xi}_j = \left(\lambda_{min}+(j_{\lambda}+\frac{1}{2})\Delta{\lambda}, {\phi}_{min}+(j_{\phi}+\frac{1}{2})\Delta{\phi}, -{h}_{min}-(j_h+\frac{1}{2})\Delta{h} \right),
    \end{equation}               
    where $j=j_{\lambda} + j_{\phi}\times n_{\lambda} + j_h \times (n_{\lambda} \times n_{\phi}), j_{\lambda} \in \mathbb{Z}[0,n_{\lambda}-1], j_{\phi} \in \mathbb{Z}[0,n_{\phi}-1],$ and $j_h \in \mathbb{Z}[0,n_h-1].$
    After calculating the center coordinates in ([deg], [deg], [km]), we can easily transform them to the geocentric rectangular coordinates using the same formulae in Eqs. (\ref{eq:earth-centered_x} -- \ref{eq:earth-centered_z}). Whenever using Eq. (\ref{eq:earth-centered_z}), the outward normal is used for the positive sign of the depth. 
    Another important quantity about the reference volume is the actual volume of individual reference volume element. In view of the curved ellipsoidal lithosphere, the volume of the $j$th reference volume element $V_j$ [$\text{km}^3$] is calculated by 
    \begin{equation}\label{eq:del_vol}
    V_j = 4\pi \|\bm{\xi}_j \|^2_2 \Delta h \frac{\Delta \lambda}{360^\circ} \frac{\Delta \phi}{180^\circ}
    \end{equation}              
    }
\end{enumerate}
\textbf{Discretization of convolved information index. }For the integration over a discrete 3D point cloud, with an uniformity assumption over a reference volume, the approximation is given by  
\begin{equation}\label{eq:nonlocal_discrete}
\overline{II}_S^{(t)}(\bm{\xi}_{j}; L_k) \cong  \sum_{i=1}^{n^{(t)}} { \omega(\bm{\xi}_j, \textbf{x}_{i}^{(t)}; L_k) II_{local}^{(t)}(\textbf{x}_{i}^{(t)}) V(\textbf{x}_{i}^{(t)})}
\end{equation}
where $V(\textbf{x}_{i}^{(t)})$ is the volume which contains the $i$th hypocenter in the lithosphere domain at time $(t)$ and is assumed to be 1 $ \text{km}^3$. Rationales behind this unit volume assignment to each hypocenter point's $V(\textbf{x}_{i}^{(t)})$ are twofold. First, the spatial impact of each hypocenter is already taken into account by the $II_{local}$'s weighted spatial integration in Eqs. (\ref{eq:nonlocal}) and (\ref{eq:nonlocal_discrete}). Second, each epoch has new hypocenters emerging at the random locations and with a random total count. To retain the physical consistency of the II, i.e. the more earthquakes the more released strain energy, it is efficient to use the unit volume per hypocenter. Unlike $V(\textbf{x}_{i}^{(t)})$, the $j$th the reference volume does have actual spatial volume (denoted as $V_j$) which is accurately calculated on the Earth ellipsoid reference (see the algorithm in corresponding section in \cite{Supplementary}). 
For the discretization of the integration over the time space, 
\begin{equation}\label{eq:nonlocal_discrete-spatio-temporal}
\overline{II}_{ST}^{(t)}(\bm{\xi}_{j}; L_k, T_l) \cong  \sum_{i=1}^{n_{epoch}} { \omega(\tau_i; T_l) \overline{II}_S^{(t_i)}(\bm{\xi}_{j}; L_k) {\Delta t}_i}
\end{equation}
where $\tau_i = |t - t_i|, t \geq t_i$ and $\Delta t_i$ is the incremental time lapse over one epoch, which is assumed to be constant $\Delta t_i = 1$ epoch (i.e. one month). $T_l \in \mathbb{R}^{+}$, $l = 1,..., n_T$. With a larger value of $T_l$, the earthquake events across a longer past period can be incorporated at the expense of over-smoothing effect; with a smaller $T_l$, a higher priority on the most recent earthquakes to the present time at the expense of local spikes or over-fitting effect. Choosing values and total counts of $T_l$ is subject to learning and prediction accuracy. To some extent, this inclusion of temporal effects by ML shares the similar notions of the well-known long short-term memory (LSTM) and its variants \cite{Hochreiter:1997, Gers:2000} that uses selective remembering and forgetting in the time axis.
%
\newline
\textbf{Holistic error measure specialized for the earthquake prediction. }
Another notable challenge of this study is rooted in the question of how to define an effective and efficient earthquake-specialized error (equivalently, loss or fitness). Sufficiently large earthquakes (e.g. moment magnitude $>$ 4.0) are not regarded as a point-wise phenomenon (e.g., rupture area $>$ 1 km$^2$; \cite{Yang:2012, Kanamori:1975}), but rather they span certain spaces. A successful error measure for ML should be able to holistically quantify the discrepancy in magnitudes, locations, and false warnings of the predicted earthquakes. To fulfill such multifaceted objectives, this study proposed a comprehensive error measure. As explained in Fig. \ref{fig:Mag_Dist_Error}, a good prediction rule should be able to predict not only the magnitudes of future events but also their locations in the three-dimensional lithosphere with the smallest number of false predictions. Still, this study's error measure is open to improvement by incorporating further computational schemes and/or deeper physics-ingrained terms.     
In particular, this study proposes a holistic error function $\mathcal{J}$ in Eq. (\ref{eq:cost_type4}) that accommodates diverse errors in magnitude, location and false alarms. The proposed error function places higher importance on the large events. 
\begin{equation}\label{eq:cost_type4}
\mathcal{J}(s) = (1-a_{cnt}) \sum_{{\tilde{k}} \in Top} \omega_{MD}^{(\tilde{k})} E_{MD}^{({\tilde{k}})}/n(Top) + a_{cnt} E_{cnt}
\end{equation}
\begin{equation}\label{eq:cost_mag_distance}
E_{MD}^{({\tilde{k}})}:=a_M\text{erf}\left( \frac{|M_{obs}^{(t+1)}(\bm{\xi}_{\tilde{k}}) - M_{pred}^{(t+1)}(\bm{\xi}_{k^*})|}{M_{obs}^{(t+1)}({\bm{\xi}}_{\tilde{k}})} \right) + (1-a_M)\text{erf}\left(\frac{||{\bm{\xi}_{\tilde{k}}}-\bm{\xi}_{k^*}||_2}{r_{max}} \right)
\end{equation}
\begin{equation}\label{eq:cost_wrong_count_option3}
E_{cnt}:=  \frac{1}{2}\text{erf}\left(\frac{|n(Top)-n(Top_{pred})|}{n(Top)} \right) + \frac{1}{2}\text{erf}\left(\sum_{\forall\tilde{k} \in Top_{pred}^{-1}}\frac{|M_{thr} - M_{pred}^{(t+1)}(\bm{\xi}_{\tilde{k}})|/M_{thr}}{n(Top_{pred}^{-1})} \right)
\end{equation}
After sorting the observed real magnitudes in descending order, we can obtain $Top$, a set of indices of reference volumes that contains the sorted real magnitudes greater than $M_{thr}$, $Top := \{{\tilde{k}} \; \text{of} \: {\bm{\xi}_{\tilde{k}}}| \; M_{obs}^{(t+1)}({\bm{\xi}_{\tilde1}})\geq M_{obs}^{(t+1)}({\bm{\xi}_{\tilde2}})\geq \cdot \cdot \cdot > M_{thr}\}$. Similarly, we obtain $Top_{pred}$, a set of indices of reference volumes that contains the sorted predicted magnitudes, $Top_{pred} := \{{\tilde{k}} \; \text{of} \: {\bm{\xi}_{\tilde{k}}}| \; M_{pred}^{(t+1)}({\bm{\xi}_{\tilde1}})\geq M_{pred}^{(t+1)}({\bm{\xi}_{\tilde2}}) \cdot \cdot \cdot > M_{thr}\}$. 
${k^*}({\tilde{k}}) \in Top_{pred}$ means the index of the spatially closest reference volume to ${\tilde{k}} \in Top$, obtained by ${k^*}({\tilde{k}}) := \text{argmin}_{\forall k \in {Top_{pred}}}||{\bm{\xi}_{\tilde{k}}}-\bm{\xi}_{k}||_2$. 
$E_{MD}^{(k)}$ in Eq. (\ref{eq:cost_mag_distance}) considers a weighted average of errors in magnitude and location of earthquakes larger than the threshold $M_{thr}$. $\text{erf(.)}$ is the Gauss error function used for mapping real-valued error $\in \mathbb{R}[0,\infty)$ to the range of [0,1] and its general definition is given by $\text{erf(z)} := \frac{2}{\sqrt\pi}\int_{0}^{z} e^{-t^2}dt$ where $z \in \mathbb{C}$ and $\text{erf}(z) \in [-1,1]$. 
A weight coefficient $a_M\in \mathbb{R}[0,1]$ determines the relative importance of magnitude error compared to the location error. Here, $a_M=0.5$ is used meaning the same importance in predicting magnitudes and locations. The proposed error function addresses the correct and false prediction of total events' count above the threshold by the second term of the right hand side of Eq. (\ref{eq:cost_type4}). The relative importance of the counting error is weighted by $a_{cnt}\in\mathbb{R}[0,1]$, herein $a_{cnt}=0.1$ is used. 
%
$r_{max} = 200$ km is used to normalize the distance prediction error. Predictions only within $r_{max}$ are considered in the error calculation while the predictions beyond $r_{max}$ are regarded as incorrect predictions. Also, the proposed error measure puts increasing weights on the larger real earthquakes through $\omega_{MD}^{(\tilde{k})}$ defined as $\omega_{MD}^{(\tilde{k})}:=\text{exp}(M_{obs}^{(t+1)}(\bm{\xi}_{\tilde{k}})/10.0); \: \tilde{k} \in Top$, which helps improve the accuracy of predicting larger, rare events. The additional error term $E_{cnt}$ of Eq. (\ref{eq:cost_wrong_count_option3}) quantifies the wrong predictions since such ``false alarm'' may hamper reliability of the prediction and result in substantial societal cost. $E_{cnt}$ consists of two terms, the first term is about how many false alarms happened in terms of the total count while the second term is about how far the false alarms deviate from the minimum threshold $M_{thr}$. In Eq. (\ref{eq:cost_wrong_count_option3}) $Top_{pred}^{-1}:= \{\tilde{k}  \; | {\tilde{k}\in Top_{pred}}$ \& never used in $E_{MD}^{({\tilde{k}})} \}.$
%
\newline
\textbf{Data availability. }The data supporting the plots and other findings of this study
are available from the corresponding author upon reasonable request.

\section*{Acknowledgments}
This work was supported by the National Science Foundation under grants CSSI-1931380 (in part). The research reported in this paper is partially supported by the HPC@ISU equipment at Iowa State University, some of which has been purchased through funding provided by the NSF under MRI grant CNS-1229081 and CRI grant 1205413. 
\section*{{Author contributions}} 
Cho is responsible for all algorithms and programs presented as well as writing of the manuscript. 
\section*{Additional Information}
{\bf{Supplementary Information}} accompanies this paper at \cite{Supplementary}.
\newline
{\bf{Competing interests:}} The author declares no competing interests.  
\newpage{}
{\centering
{\Large\textbf{Initial Foundation for Predicting Individual Earthquake's Location and Magnitude by Using Glass-Box Physics Rule Learner}} \\
\textbf{In Ho Cho$^{1}$} \\
\textbf{$^{1}$CCEE Department, Iowa State University, Ames, IA 50011, USA}
}
\newline
%
%
%
\captionsetup[table]{labelfont={bf},labelformat={default},labelsep=period,name={Table}}
\renewcommand{\thetable}{S\arabic{table}}

\setcounter{figure}{0}
\renewcommand{\thefigure}{S\arabic{figure}}
%
%

%
%
%
%
\subsection*{Limits of direct use of existing machine learning methods }
Despite many triumphs of recent ML methods \cite{Brown:2018, Mnih:2015}, the direct adoption of the existing ML methods for this study's goal is not promising for several reasons. This study needs to explore multifaceted physical data sets defined over multiple-dimensions and seeks to unravel generic expressions (i.e. glass-box learning) in lieu of final predictions (i.e. black-box learning). Pursuing hidden rules, this study is aligned with the so-called physics-guided ML paradigms in broad science and engineering \cite{Karpatne_et_al:2017, Cho:2019, Raissi:2020, Champion_et_al:2019}. Recently, the author applied the glass-box physics rule learner (GPRL) to hidden physics of nano-scale tribocharging phenomena \cite{Cho:2020}. This study inherits the central notions of the GPRL framework and substantially expand it for the hidden rules of imminent earthquake prediction. Such extensions of GPRL is possible since it can deal with multifaceted measurements over higher dimensions, infuse basic physics and scientists knowledge, extract hidden rule's generic expressions, and evolve the rules with increasing data through the Bayesian update scheme.   
\begin{figure}[h]
  \centering
  \includegraphics[width=0.50\textwidth]{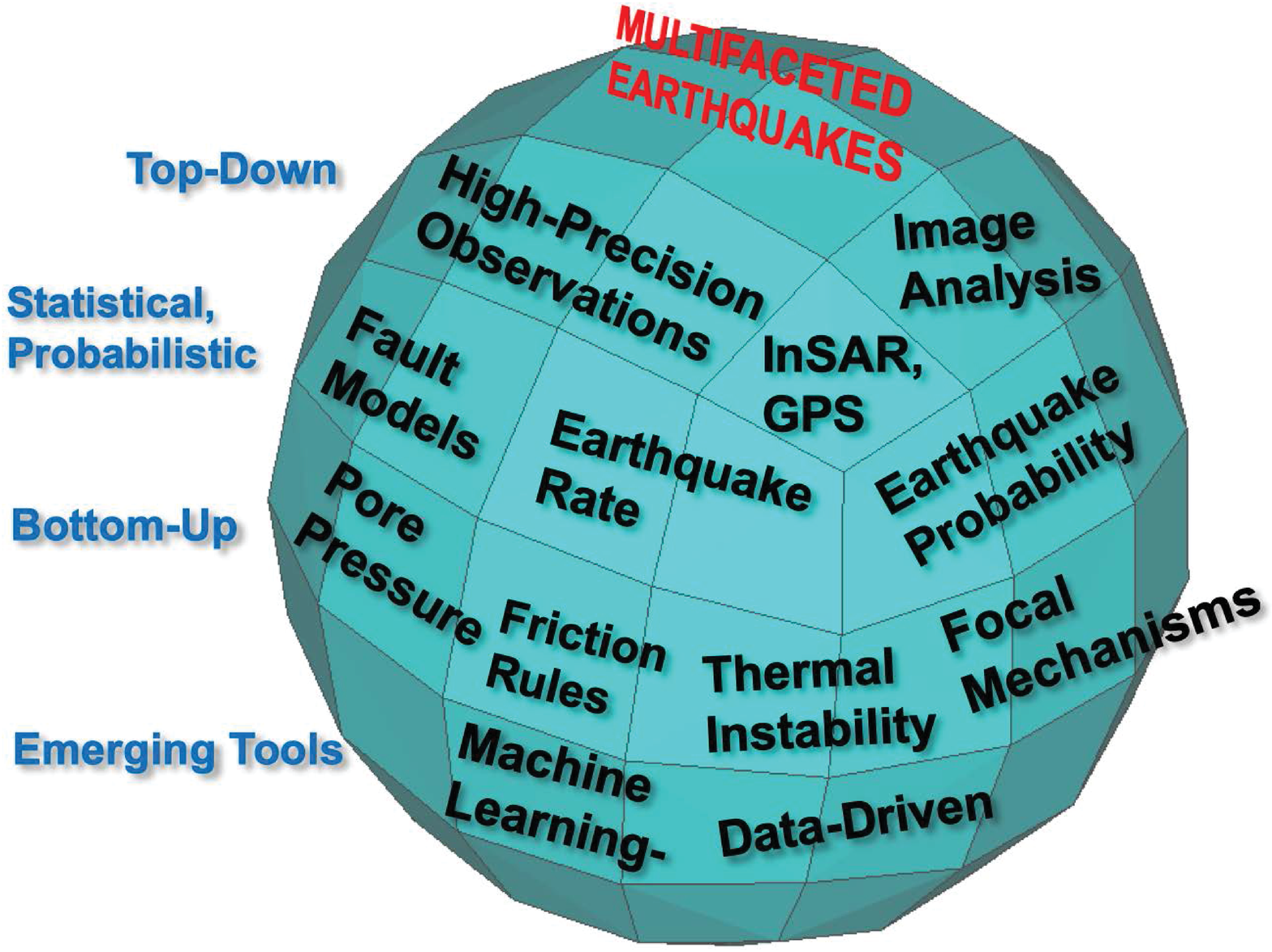}
  \caption{\textbf{Multifaceted approaches to Earthquakes.} }
  \label{fig:Multifaceted_Physics}
\end{figure}
\begin{figure}[h]
  \centering
  \includegraphics[width=1.0\textwidth]{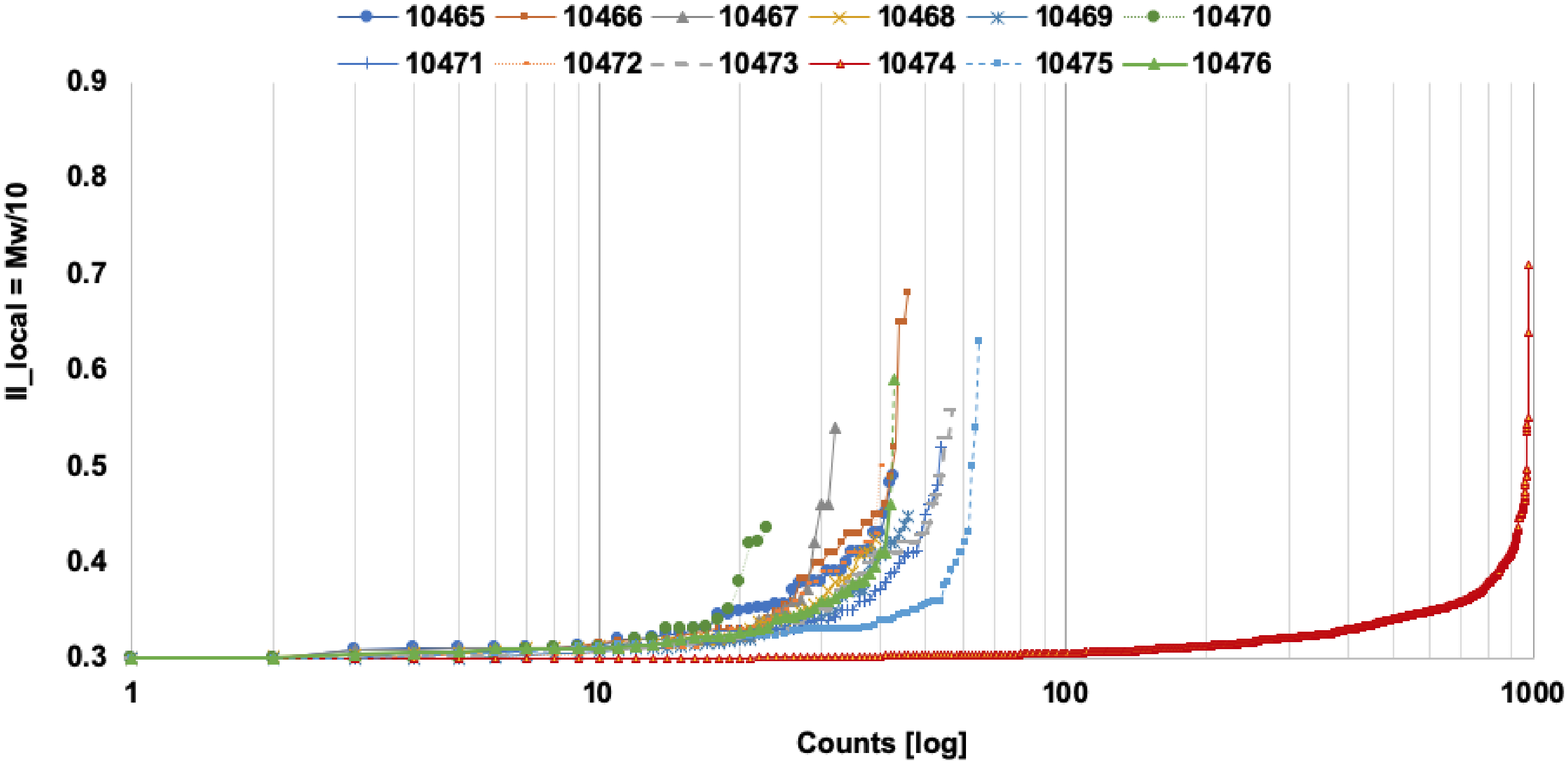}
  \caption{\textbf{Point-wise information index, $II_{local}.$} Events above 0.3 (i.e. $M_w \geq 3.0$) are shown. Each line corresponds to the events during one-month period from October 2018 (epoch number 10465) to September 2019 (10476).}
  \label{fig:Local_II_from10465}
\end{figure}
\begin{figure}[h]
  \centering
  \includegraphics[width=1.0\textwidth]{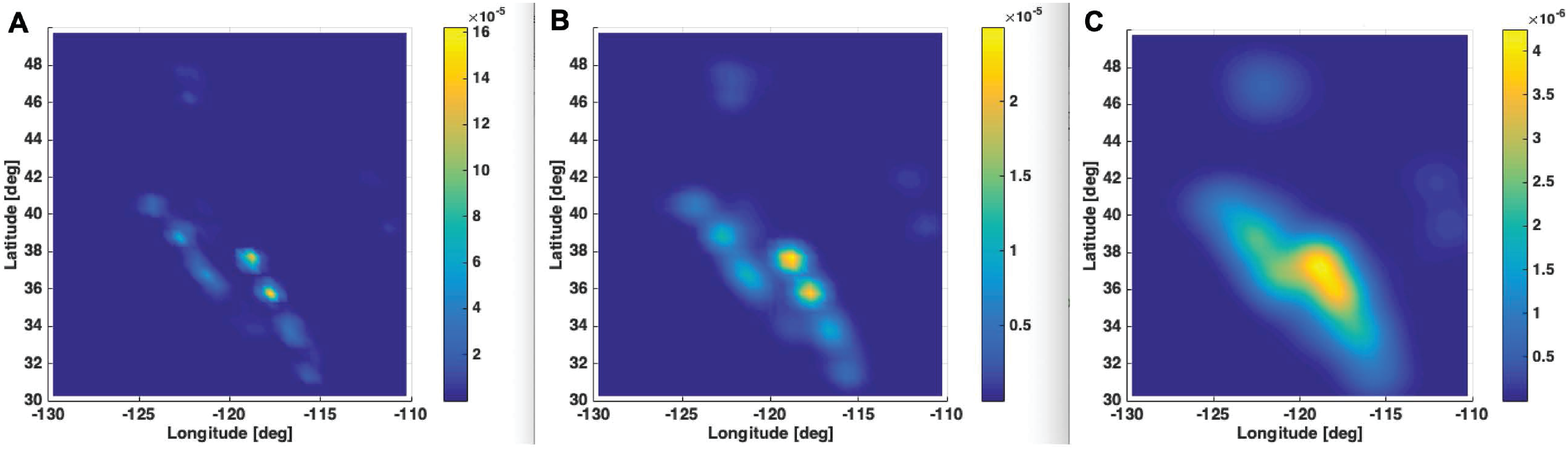}
  \caption{\textbf{Convolved Spatial II at the depth z = 12.5 km: (A)} Calculated with $L=25$ km. \textbf{(B)} With $L=50$ km. \textbf{(C)} $L=100$ km.}
  \label{fig:Spatial_II_z12_5km}
\end{figure}
%
%
\begin{table}[b]
  \caption{Earthquake catalog and the associated training epochs [1 month = 1 epoch]}
  \label{table:Epoch_list}
  \centering
  \begin{tabular}{llll}
    \hline
    Year (begin)     & Year (end) & Epoch (begin) & Epoch (end)\\
    \hline
    1980  &  1984 & 10000  & 10059     \\
    1985  &  1989 & 10060  & 10119     \\
    1990  &  1994 & 10120  & 10179     \\  
    1995  &  1999 & 10180  & 10239     \\  
    2000  &  2004 & 10240  & 10299     \\  
    2005  &  2009 & 10300  & 10359     \\  
    2010  &  2014 & 10360  & 10419     \\  
    2015  &  2019 (up to Oct.) & 10420  & 10477     \\  
   \hline
  \end{tabular}
\end{table}
\subsection*{Rescaling of information index to $\mathbb{R}$[0,1] } 
\begin{figure}
  \centering
  \includegraphics[width=1.0\textwidth]{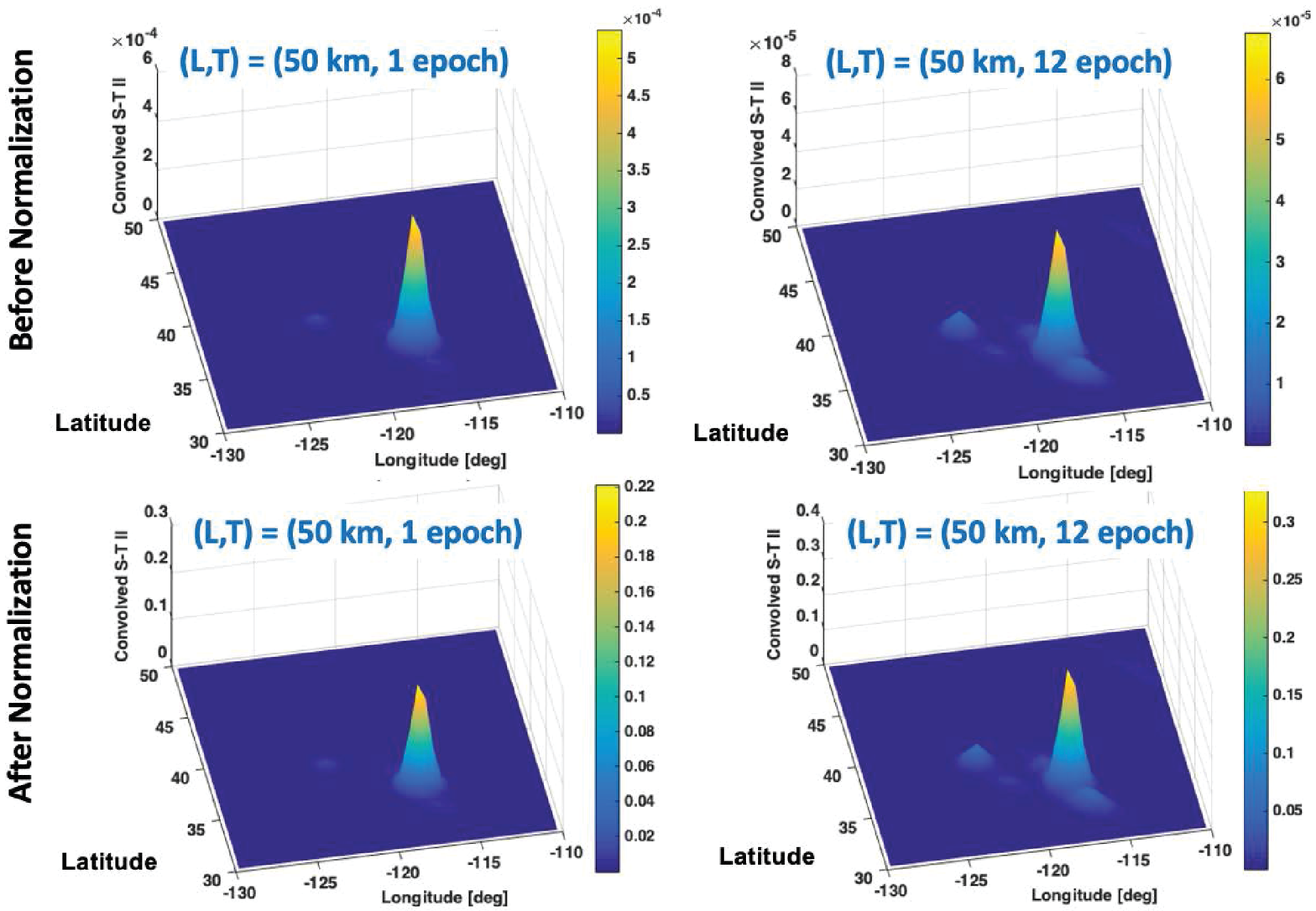}
  \caption{\textbf{Convolved Spatio-Temporal II after Normalization with the Physical Upper Bound.} The same shape but dramatic increase in the scale of values from $10^{-5}$ to the desired range of [0,1].}
  \label{fig:Normalized_II_ST}
\end{figure}

The convolved spatio-temporal II may span substantially small values which may hamper the searching power of the adopted link function (LF). For instance, the cubic regression spline (CRS)-based LF takes a covariate ranging between [0,1], and thus if the covariate is very small the subsequent learning may not be efficient. Thus, it is helpful to rescale the convolved spatio-temporal II to [0,1]. One immediate normalization would be using the maximum range of the convolved spatio-temporal II during the current epoch $(t)$ as
\begin{equation}\label{eq:Rescale-spatio-temporal}
\overline{II}_{ST}^{(t)}(\bm{\xi}_{j}; L_k, T_l) =\frac{\overline{II}_{ST}^{(t)}(\bm{\xi}_{j}; L_k, T_l) -\text{min}_{\forall \:\bm\xi_j }[{\overline{II}_{ST}^{(t)}(\bm{\xi}_{j}; L_k, T_l)}]}{|\text{max}_{\forall \:\bm\xi_j }[{\overline{II}_{ST}^{(t)}(\bm{\xi}_{j}; L_k, T_l)}] - \text{min}_{\forall \:\bm\xi_j }[{\overline{II}_{ST}^{(t)}(\bm{\xi}_{j}; L_k, T_l)}]|}
\end{equation}
However, this simple normalization may not guarantee the consistency over different epochs since each epoch may have different ranges of the spatio-temporal IIs. Therefore, it is more robust to use the normalization with the upper bound (see a brief proof in the following section) of the spatio-temporal II as
\begin{equation}\label{eq:Rescale-spatio-temporal_with_UpperBound}
\overline{II}_{ST}^{(t)}(\bm{\xi}_{j}; L_k, T_l) =
\frac{\overline{II}_{ST}^{(t)}(\bm{\xi}_{j}; L_k, T_l)}
{{\widetilde{n}_{epoch}}(T_l (2\pi)^{1/2})^{-1}\times {\widetilde{n}^{(t)}}(L_k (2\pi)^{1/2})^{-3}}
\end{equation}
where it is reasonably assumed that $\widetilde{n}_{epoch} = 60$ epochs (i.e., 5 years); $\widetilde{n}^{(t)} = 200 $ (i.e. at most 200 events larger than magnitude 3.0 per month) based on the past earthquakes in the catalog database \cite{USGS_Catalog}. This assumption of constant $\widetilde{n}_{epoch}$ and $\widetilde{n}^{(t)}$ helps the spatio-temporal II be consistently normalized for a given pair of $(L_k, T_l)$. As long as the consistency is being held and the resulting II is within [0,1], these values of $\widetilde{n}_{epoch}$ and $\widetilde{n}^{(t)}$ can be changed by researchers for the upper bound-based normalization. Fig. \ref{fig:Normalized_II_ST} presents the positive role of the upper bound-based normalization that boosts the convolved spatio-temporal II to the manageable range of [0,1].  
\begin{figure}
  \centering
  \includegraphics[width=1.0\textwidth]{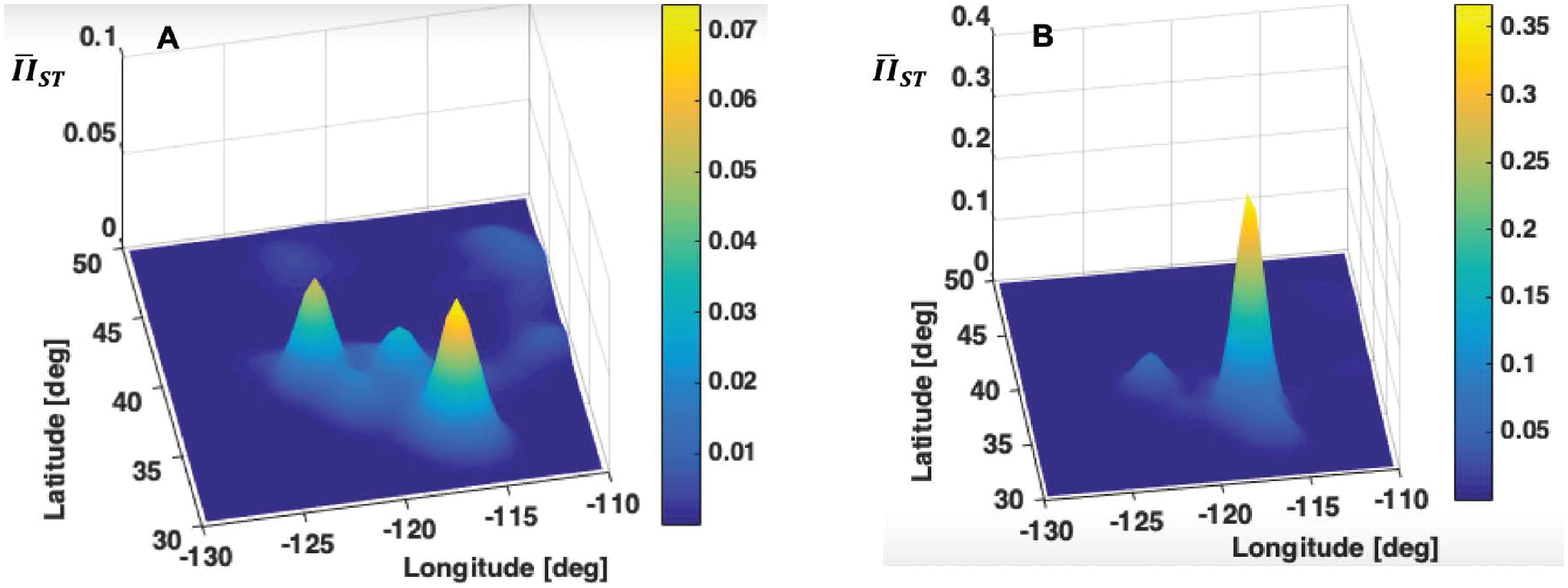}
  \caption{\textbf{Comparison of two convolved spatio-temporal II: (A) } $\overline{II}_{ST}$ over 12 epochs from 10462 to 10472 showing relatively low seismic activity; \textbf{(B) } $\overline{II}_{ST}$ over 12 epochs from 10465 to 10475 showing relatively high seismic activity with peaks five fold larger than \textbf{(A)}, mainly due to the notably active epoch 10474 (see Fig. \ref{fig:Local_II_from10465} and Fig. \ref{fig:Convolved_II_Definition}). All $\overline{II}_{ST}$s are generated with $(L,T) = $(75 km, 12 epochs) and the plot depth is $z = 12.5$ km.}
  \label{fig:II_ST_12epochs_from10462_or_10465}
\end{figure}
To consider the upper bound of the spatio-temporal II, it is necessary to consider the upper bound of the convolved spatial II. In lieu of exploring all possible ranges, bound-aware searching helps fast searching, and thus the derivation of the upper bound is presented in the following section.
\subsection*{Upper bound of spatio-temporal information index}\label{sec:proof_upper_bound}
To consider the upper bound of the spatio-temporal II, it is necessary to consider the upper bound of the convolved spatial II.
\begin{equation}\label{eq:nonlocal_discrete_proof}
\overline{II}_S^{(t)}(\bm{\xi}_{j}; L_k) \cong  \sum_{i=1}^{n^{(t)}} { \omega(\bm{\xi}_j, \textbf{x}_{i}^{(t)}; L_k) II_{local}^{(t)}(\textbf{x}_{i}^{(t)}) V(\textbf{x}_{i}^{(t)})}
\end{equation}
\begin{align}\label{eq:nonlocal_discrete_proof2}
    \leq \sum_{i=1}^{n^{(t)}} { \omega(\bm{\xi}_j, \textbf{x}_{i}^{(t)}; L_k) \times 1.0 \times 1.0}   
\end{align}
\begin{align}\label{eq:nonlocal_discrete_proof3}
    \leq n^{(t)} \times { \omega(\bm{\xi}_j =\textbf{x}_{i}^{(t)}; L_k)} = n^{(t)} \times (L_k(2\pi)^{1/2})^{-3}  
\end{align}
Here, the inequality Eq. \ref{eq:nonlocal_discrete_proof2} assumes the extreme case when all events are maximum magnitude of 10 and thus $II_{local}=10/10=1.0$. $V(\textbf{x}_{i}^{(t)})=1.0$ is physically explained in the text around Eq. \ref{eq:nonlocal_discrete}. The last inequality Eq. \ref{eq:nonlocal_discrete_proof3} assumes another extreme case when all events happen at the $j$th reference volume and all the Gaussian weights take the maximum value. Thus, the maximum physically conceivable upper bound of $\overline{II}_S^{(t)}(\bm{\xi}_{j}; L_k)$ is $n^{(t)} (L_k(2\pi)^{1/2})^{-3}$.
Now we can consider the upper bound of the convolved spatio-temporal II. 
\begin{equation}\label{eq:nonlocal_discrete-spatio-temporal_proof}
\overline{II}_{ST}^{(t)}(\bm{\xi}_{j}; L_k, T_l) \cong  \sum_{i=1}^{n_{epoch}} { \omega(\tau_i; T_l) \overline{II}_S^{(t_i)}(\bm{\xi}_{j}; L_k) {\Delta t}_i}
\end{equation}
\begin{align}\label{eq:nonlocal_discrete-spatio-temporal_proof2}
    \leq \sum_{i=1}^{n_{epoch}} { \omega(\tau_i; T_l) \left(n^{(t_i)} (L_k(2\pi)^{1/2})^{-3}\right) {\Delta t}_i}
\end{align}
\begin{align}\label{eq:nonlocal_discrete-spatio-temporal_proof3}
    \leq n_{epoch} \times { \omega(\tau_i=0; T_l) \left(n^{(t_i)} (L_k(2\pi)^{1/2})^{-3}\right) {\Delta t}_i}
\end{align}
\begin{align}\label{eq:nonlocal_discrete-spatio-temporal_proof4}
    \leq (n_{epoch} \times  (T_l(2\pi)^{1/2})^{-1}) \left(n^{(t_i)} (L_k(2\pi)^{1/2})^{-3}\right) {\Delta t}_i
\end{align}
The inequality Eq. \ref{eq:nonlocal_discrete-spatio-temporal_proof3} assumes the extreme case that all past earthquake events took place current epoch and thus gives $\tau_i=0$. As explained before, $\forall{\Delta t}_i=1$ epoch which means 1 month in this study. Therefore, the physically conceivable upper bound of the convolved spatio-temporal II is $\left(n_{epoch} (T_l(2\pi)^{1/2})^{-1}\right) \left(n^{(t_i)} (L_k(2\pi)^{1/2})^{-3}\right)$.
%
\subsection*{Training with an epoch frame}
\begin{figure}
  \centering
  \includegraphics[width=1.0\textwidth]{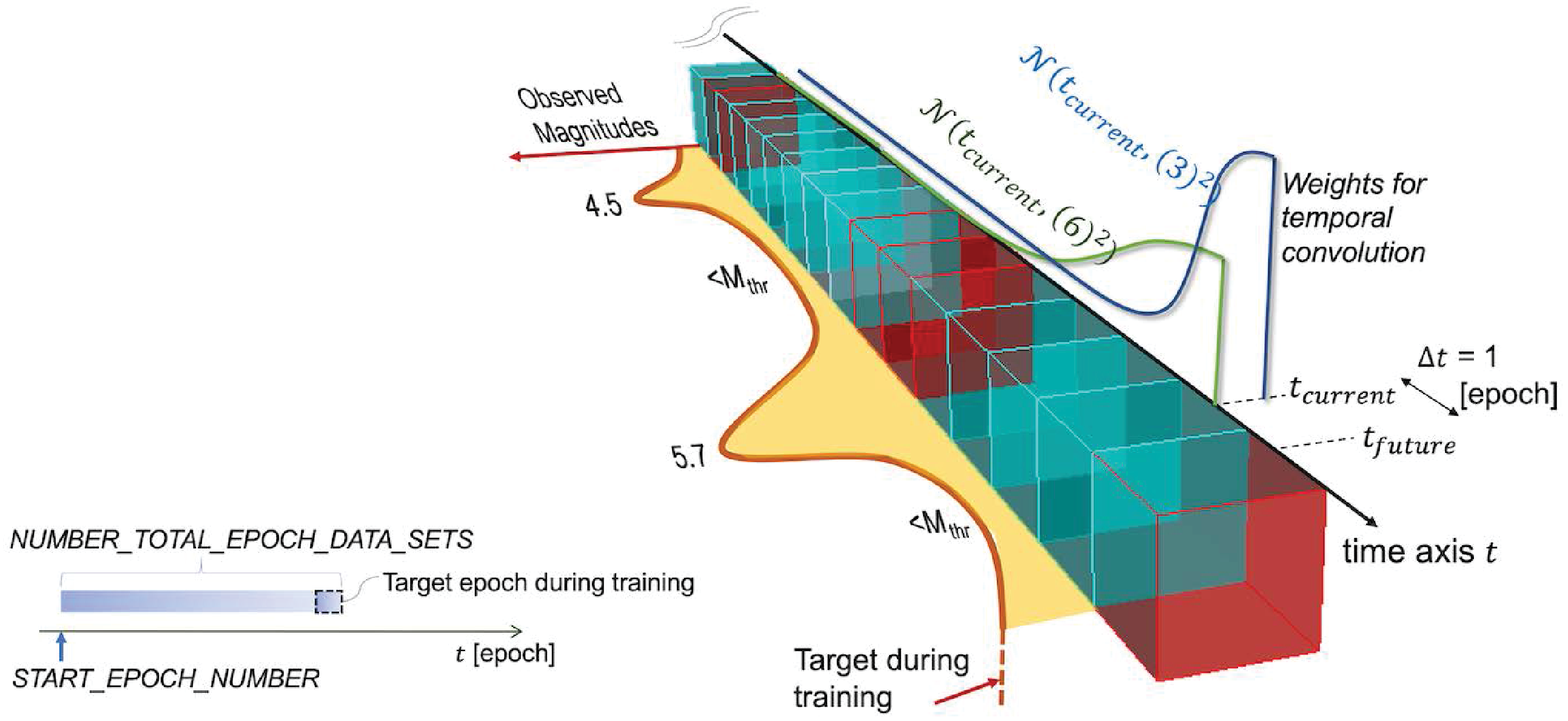}
  \caption{\textbf{Illustration of training with a epoch frames.} The last epoch of the time frame is used for target during training. Between different time frames, the Bayesian inheritance of the prior-best rules may take place.}
  \label{fig:Temporal_Convolution_Illustration2}
\end{figure}

Fig. \ref{fig:Temporal_Convolution_Illustration2} explains how training takes place with a time frame. Departing from the first epoch (marked by ``\textit{START\_EPOCH\_NUMBER}") with the length of ``\textit{NUMBER\_TOTAL\_EPOCH\_DATA\_SETS}", all the convolved spatio-temporal IIs within the epoch frame are used for training and then the identified rule is used to predict the earthquakes of the last epoch of the frame (marked by dashed box). After training, the best-so-far rules are identified. For instance, when ``\textit{START\_EPOCH\_NUMBER} = 10462", and ``\textit{NUMBER\_TOTAL\_EPOCH\_DATA\_SETS} = 12", the training is conducted with $\forall \overline{II}_{ST}^{(t)}, t=[10462, 10472]$ (in total 11 epochs) in order to identify the hidden rules that can best reproduce earthquakes happening at the last epoch 10473. These identified rules are stored as a prior generation. Between different time frames, the prior best generation of the identified rules can be inherited via the combination of Bayesian update and evolutionary algorithm.
\section*{Flexible and expressive link functions}\label{sec:link}
Place top priority on the interpretability, this study proposes to adopt an expressive link function (LF) using transparent, flexible basis that is capable of describing a mathematical expression between the convolved spatio-temporal II, $\overline{II}_{ST}$ and the hidden physical rules. LF is denoted as $\mathcal{L}(\overline{II}_{ST}; \bm{\uptheta})$ where $\bm{\uptheta}$ is a set of free parameters prescribing the LF. This study used an evolutionary algorithm coupled with the Bayesian update scheme to enable LF to continue to learn, train, and evolve. There is little restriction of choice of other forms of LFs. 
For balancing the  efficiency and interpretability, this study chose the cubic spline basis owing to its high smoothness and flexibility. The cubic spline curves consist of a few cubic polynomials connected at knots so that the curves are continuous up to the second derivatives \cite{Wood:2006}. If practical cubic spline bases \cite{Gu:2013} (denoted as $b_i$) are adopted, LFs are given as 
\begin{equation}\label{eq:link_CRS}
\mathcal{L}(\overline{II}_{ST};~\textbf{a},\textbf{x}^*) = \sum_i^p{a_{i} b_{i}(\overline{II}_{ST})}
\end{equation}
\begin{equation}\label{eq:link_CRS2}
\textbf{CRS-Based LF: }\mathcal{L}^{(k,l)}(\overline{II}_{ST}^{(t)}(\bm{\xi}_{j}; L_k, T_l) ; ~\bm{\uptheta}^{(k,l)}) = \sum_{i=1}^p{a_{i}^{(k,l)} b_{i}^{(k,l)}(\overline{II}_{ST}^{(t)}(\bm{\xi}_{j}; L_k, T_l))};
\end{equation}
where $b_{1}(x) = 1, b_{2}(x) = x,$ and 
\begin{equation}\label{eq:cubic_spline}
b_{i+2}(x) = \frac{[(x_i^* - \frac{1}{2})^2 - \frac{1}{12}][(x-\frac{1}{2})^2 - \frac{1}{12}]}{4} - \frac{[(|x-x_i^*|-\frac{1}{2})^4 - \frac{1}{2}(|x-x_i^*| - \frac{1}{2})^2 + \frac{7}{240}]}{24}, 
\end{equation}
for $i=1...p-2.$ Here, $x_i^*$ is $i_{th}$ knot location. To fully describe one LF, we need to identify $p + (p-2)$ unknowns, i.e. $\textbf{a} = \{a_1,...,a_p\}$ and $\textbf{x}^* = \{x_1^*,...,x_{(p-2)}^*\}.$ For brevity, we denote the total unknown parameters as $\bm{\uptheta} = \{\textbf{a}, \textbf{x}^*\}$ hereafter.
The adopted cubic spline bases can accommodate a variety of relation forms, ranging from a simple monotonic rule to a highly nonlinear rule. It should be noted that the adopted cubic spline basis is not for the direct regression, but for the transparent expressions of the final rule. 
\begin{equation}\label{eq:multiple_link_exponential}
\textbf{Exponential LF: } \mathcal{L}^{(k,l)}(\overline{II}_{ST}^{(t)}(\bm{\xi}_{j}; L_k, T_l) ; ~\bm{\uptheta}^{(k,l)}) = \text{exp}\left(a^{(k,l)}\overline{II}_{ST}^{(t)}(\bm{\xi}_{j}; L_k, T_l)^{b^{(k,l)}}\right) - 1
\end{equation}
\begin{figure}
  \centering
  \includegraphics[width=0.9\textwidth]{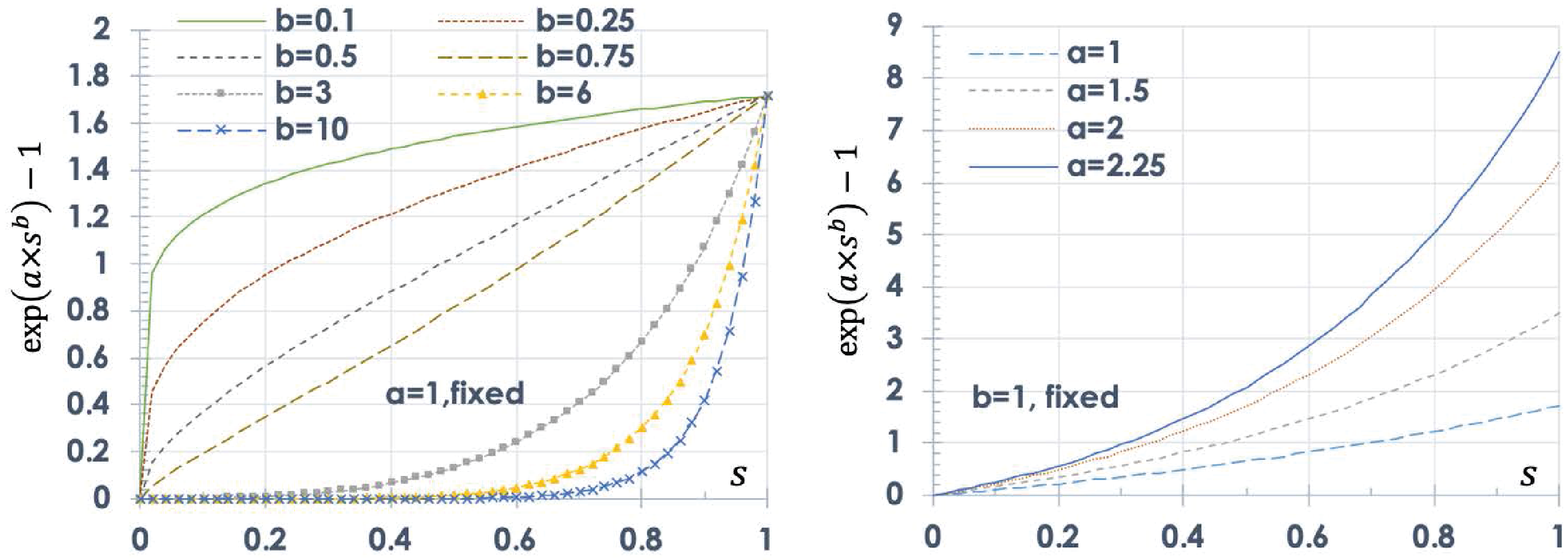}
  \caption{\textbf{Exponential LF: } The two-parameter exponential LF is adjusted by $a$ and $b$. While $a$ controls the amplitude, $b$ governs the shape.}
  \label{fig:exponential_LF}
\end{figure}
\begin{figure}
  \centering
  \includegraphics[width=0.9\textwidth]{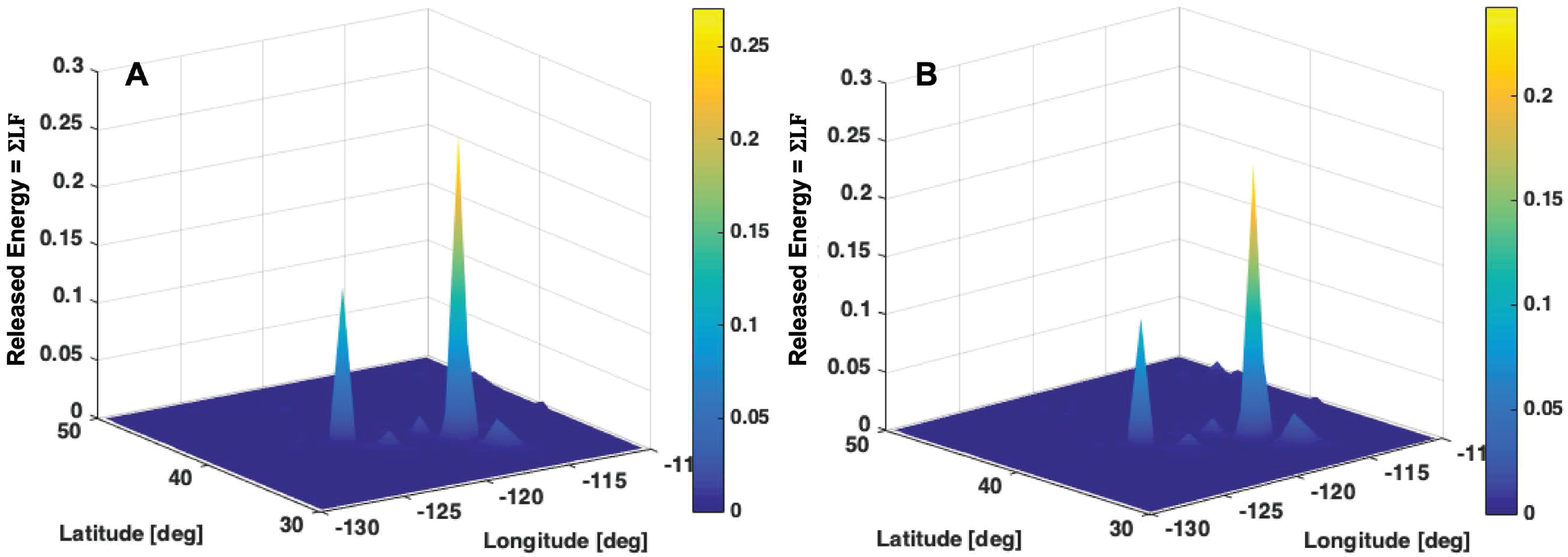}
  \caption{\textbf{Released Energy using Additive Combination of Exponential LFs: } Example plots of the released energy calculated by the additive combination of four exponential LFs with $(L,T), L=(12.5, 25)$ km and $T= (6, 12)$ epochs \textbf{(A)} at the depth $z=7.5$ km and \textbf{(B)} $z=12.5$ km.}
  \label{fig:Released_E_Exponential}
\end{figure}
%
%
%
%
\subsection*{\textbf{Time derivatives and spatial gradient of the released energy} } The time derivative of energy is physically related to the power. For later use of power term in the prediction model it is helpful to prepare time derivatives of the energy-related terms.
For the finite difference method, we adopt the backward difference scheme in view of our goal to predict the imminent earthquake in the next time epoch $(t+1)$, and thus the form is given as 
\begin{equation}\label{eq:backward_difference}
\frac{\partial \overline{II}_{ST}^{(t)(k,l)}}{\partial{t}} = \frac{ \overline{II}_{ST}^{(t)(k,l)} - \overline{II}_{ST}^{(t-1)(k,l)}}{\Delta{t}} + \mathcal{O}(\Delta t)
\end{equation}
${\partial a^{(t)(k,l)}}/{\partial{t}}$ and ${\partial b^{(t)(k,l)}}/{\partial{t}}$ are similarly calculated. 
It should be noted that the spatial gradient is with respect to the geocentric coordinate system which convey little physical and geometrical information of the earth lithosphere. Thus, it is meaningful to transform the geocetric gradient of a function $\text{f}$ to the geodetic gradient (denoted as $\nabla_{g} \text{f}$), i.e., the gradient with respect to the geodetic coordinate system ${\lambda, \phi, h}$. This can be done by multiplying Jacobian \textbf{J} as
\begin{equation}\label{eq:gradient_f_geodetic}
\nabla_{g} \text{f}(\bm\xi_j) = \textbf{J}\nabla \text{f}(\bm\xi_j),
\end{equation}
where the Jacobian's entities $\{J_{(i,j)}\}$, $i,j=1,2,3$ as given by 
\begin{equation}\label{eq:Jacobian_geodetic_1_1}
J_{(1,1)}=\frac{\partial{x}}{\partial \lambda} = -\left( DJ_1 \right)\text{cos}\phi \;\text{sin}\lambda
\end{equation}
\begin{equation}\label{eq:Jacobian_geodetic_2_1}
J_{(2,1)}=\frac{\partial{x}}{\partial \phi} = \left( DJ_2  \right)\text{cos}\phi \; \text{cos}\lambda - (DJ_1)\text{sin}\phi \; \text{cos}\lambda 
\end{equation}
\begin{equation}\label{eq:Jacobian_geodetic_3_1}
J_{(3,1)}=\frac{\partial{x}}{\partial{h}} = \text{cos}\phi \; \text{cos}\lambda 
\end{equation}
\begin{equation}\label{eq:Jacobian_geodetic_1_2}
J_{(1,2)}=\frac{\partial{y}}{\partial \lambda} = \left( DJ_1 \right)\text{cos}\phi \;\text{cos}\lambda
\end{equation}
\begin{equation}\label{eq:Jacobian_geodetic_2_2}
J_{(2,2)}=\frac{\partial{y}}{\partial \phi} = \left( DJ_2 \right)\text{cos}\phi \;\text{sin}\lambda - (DJ_1)\text{sin}\phi \; \text{sin}\lambda
\end{equation}
\begin{equation}\label{eq:Jacobian_geodetic_3_2}
J_{(3,2)}=\frac{\partial{y}}{\partial{h}} = \text{cos}\phi \; \text{sin}\lambda 
\end{equation}
\begin{equation}\label{eq:Jacobian_geodetic_1_3}
J_{(1,3)}=\frac{\partial{z}}{\partial \lambda} = 0
\end{equation}
\begin{equation}\label{eq:Jacobian_geodetic_2_3}
J_{(2,3)}=\frac{\partial{z}}{\partial \phi} = \left( DJ_4 \right)\text{sin}\phi + (DJ_3)\text{cos}\phi 
\end{equation}
\begin{equation}\label{eq:Jacobian_geodetic_3_3}
J_{(3,3)}=\frac{\partial{z}}{\partial{h}} = \text{sin}\phi 
\end{equation}
\begin{equation}\label{eq:Jacobian_geodetic_term1}
DJ_1 = \frac{a^2}{(a^2 \text{cos}^2\phi + b^2 \text{sin}^2 \phi)^{1/2}} + h 
\end{equation}
\begin{equation}\label{eq:Jacobian_geodetic_term2}
DJ_2 = \frac{a^2(a^2 \text{cos} \phi \; \text{sin} \phi - b^2\text{sin} \phi \; \text{cos} \phi)}{(a^2 \text{cos}^2\phi + b^2\text{sin}^2\phi)^{3/2}}
\end{equation}
\begin{equation}\label{eq:Jacobian_geodetic_term3}
DJ_3 = \frac{b^2}{(a^2 \text{cos}^2\phi + b^2 \text{sin}^2 \phi)^{1/2}} + h 
\end{equation}
\begin{equation}\label{eq:Jacobian_geodetic_term4}
DJ_4 = \frac{b^2(a^2 \text{cos} \phi \; \text{sin} \phi - b^2\text{sin} \phi \; \text{cos} \phi)}{(a^2 \text{cos}^2\phi + b^2\text{sin}^2\phi)^{3/2}}
\end{equation}
%
%
\begin{figure}[h]
  \centering
  \includegraphics[width=0.90\textwidth]{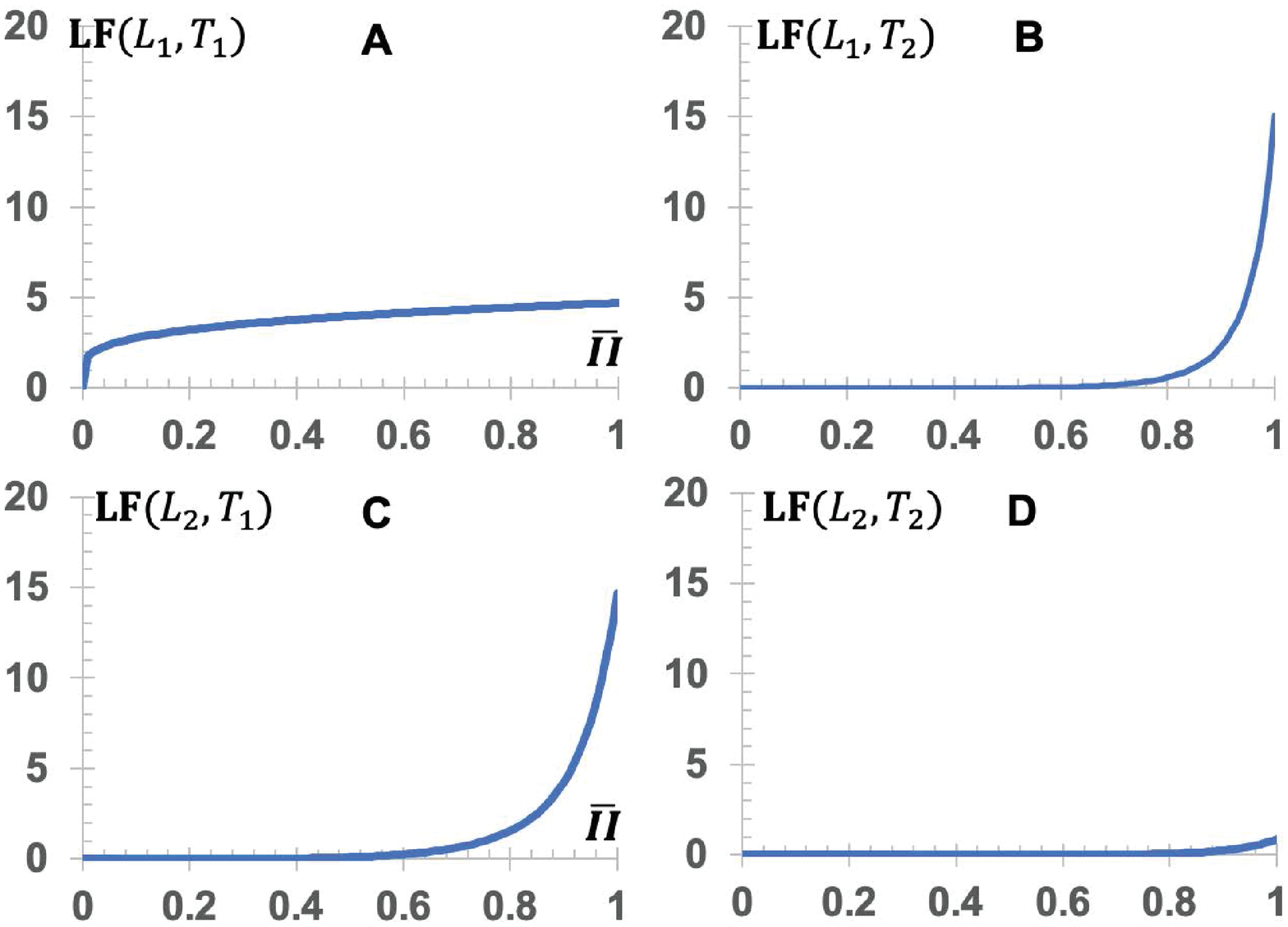}
  \caption{\textbf{Link function comparison of the identified best-so-far rule of the released energy. (A)} Two-parameter exponential LF with $(L_1, T_1)=$(10 km, 3 epochs). \textbf{(B)} $(L_1, T_1)=$(10 km, 6 epochs). \textbf{(C)} $(L_2, T_1)=$(25 km, 3 epochs). \textbf{(D)} $(L_2, T_2)=$(25 km, 6 epochs).}
  \label{fig:LF_Er_S_CA_Test48}
\end{figure}
\begin{figure}[h]
  \centering
  \includegraphics[width=1.0\textwidth]{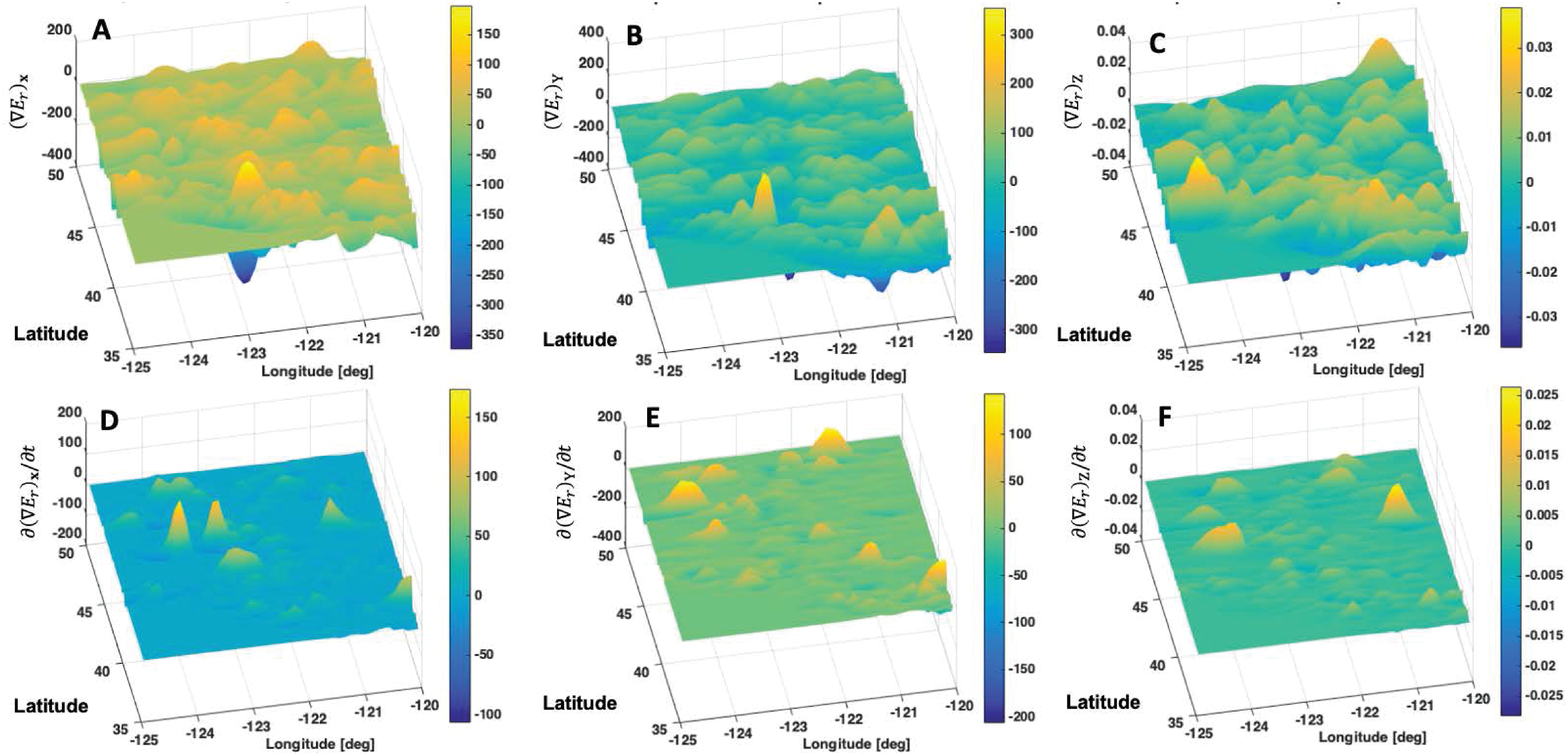}
  \caption{\textbf{Spatial gradients and time derivatives of the released energy at depth = 2.5 km generated from the past 10 years data (epochs from 10355 through 10473).} \textbf{(A-C)} At depth = 2.5 km, three components of the spatial gradients of the released energy with respect to the earth-centered XYZ system. \textbf{(D-F)} Time derivative of the spatial gradients of the released energy with the time increment of 1 epoch for derivative calculation.}
  \label{fig:PT1_z2_5km_Grad}
\end{figure}
\begin{figure}[h]
  \centering
  \includegraphics[width=1.0\textwidth]{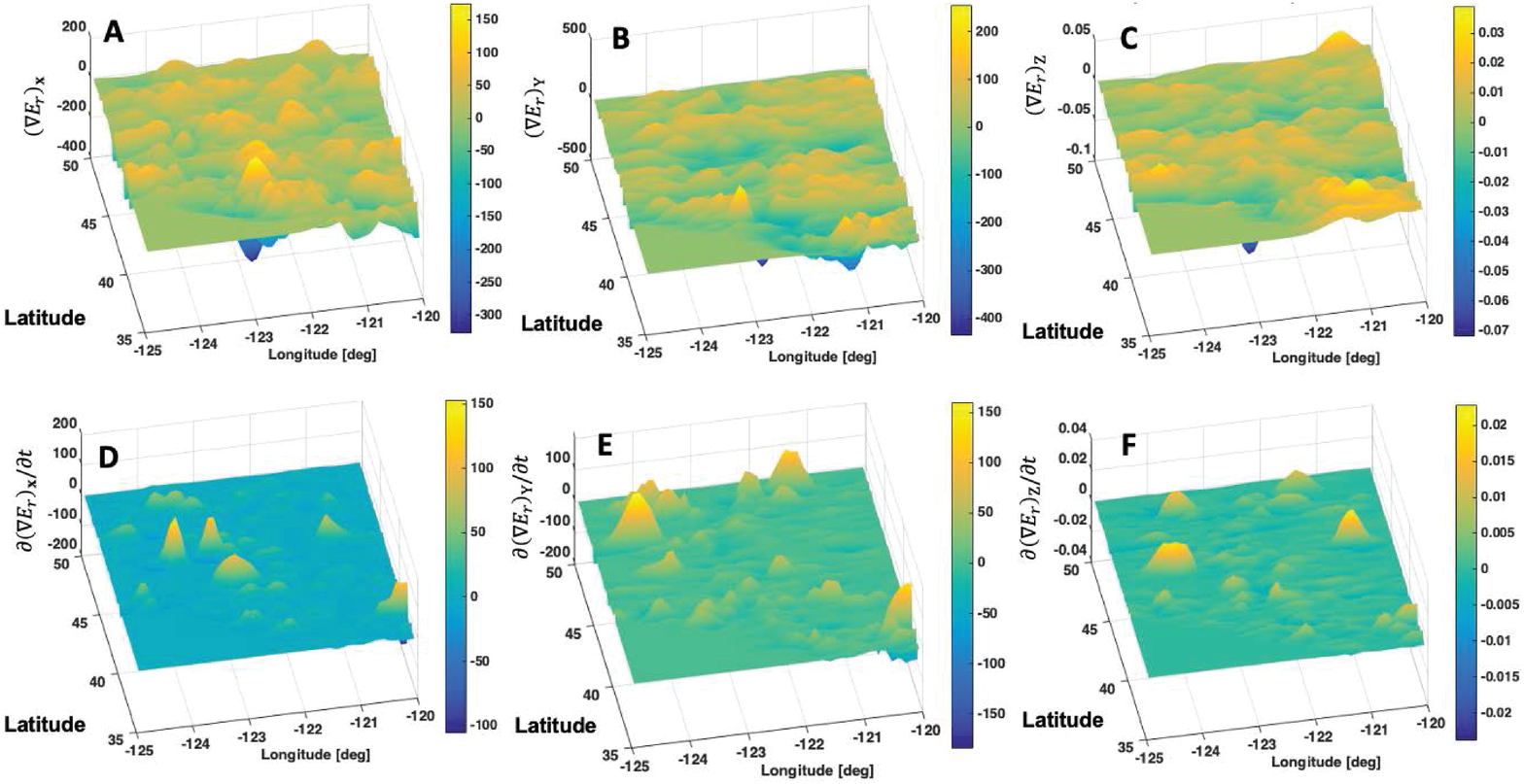}
  \caption{\textbf{Spatial gradients and time derivatives of the released energy at depth = 12.5 km generated from the past 10 years data (epochs from 10355 through 10473).} \textbf{(A-C)} Three components of the spatial gradients of the released energy with respect to the earth-centered XYZ system. \textbf{(D-F)} Time derivative of the spatial gradients of the released energy with the time increment of 1 epoch for derivative calculation.}
  \label{fig:PT1_z12_5km_Grad}
\end{figure}
\begin{figure}
  \centering
  \includegraphics[width=0.9\textwidth]{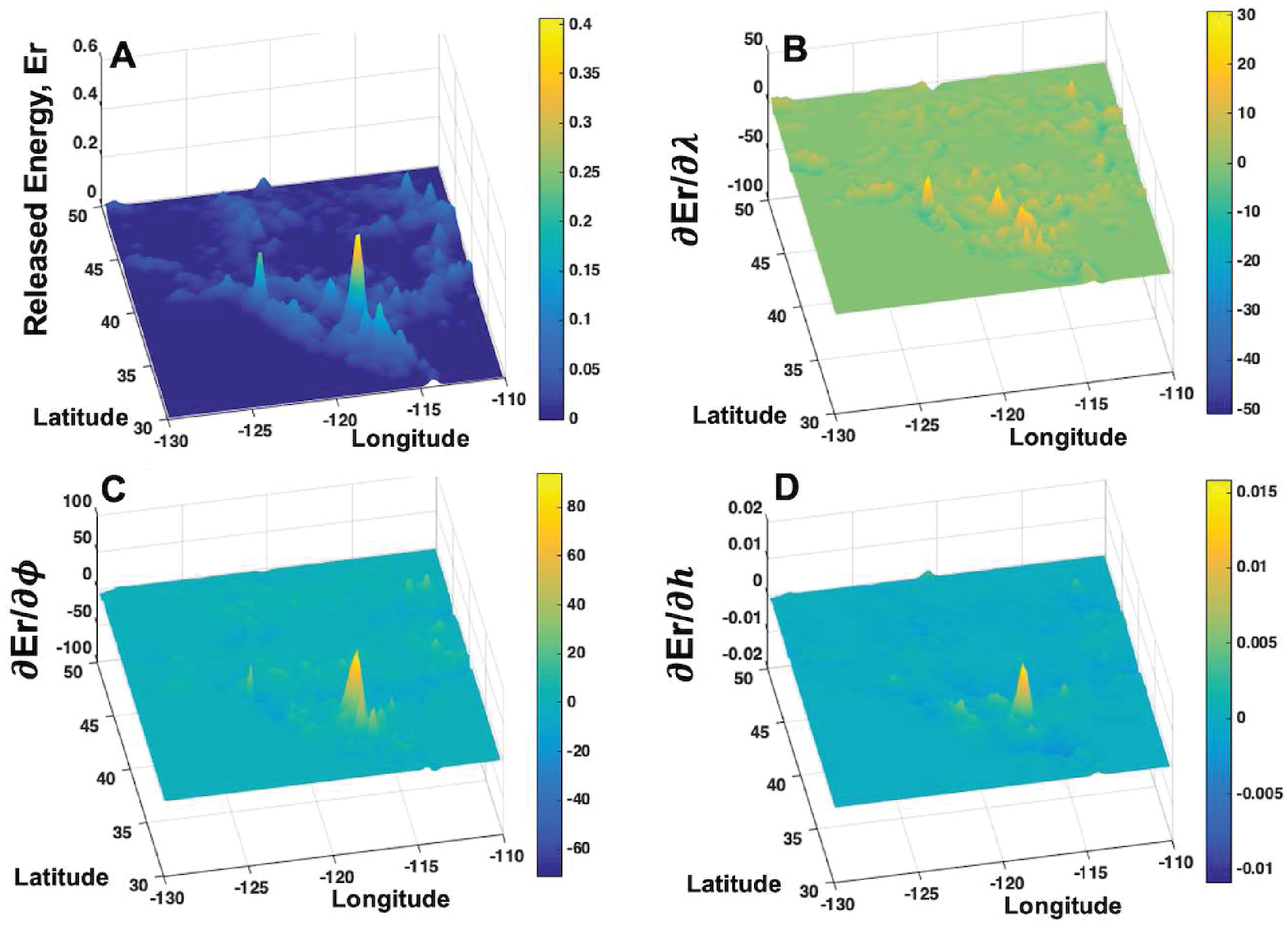}
  \caption{\textbf{Spatial Gradient of the Released Energy with respect to the Geodetic Coordinate System. (A)} The released energy generated by the multiplicative combination of four exponential LFs with $(L,T), L=(12.5, 25)$ km and $T= (6, 12)$ epochs at the depth $z=12.5$ km. \textbf{(B-D)} Spatial gradients of the released energy in the longitude $(\lambda)$, latitude $(\phi)$, and depth $(h)$ direction at depth $12.5$ km downward, respectively. The increments of (0.2 deg, 0.2 deg, 5 km) are used for spatial discretization. }
  \label{fig:Spatial_Grad_Released_E}
\end{figure}
\begin{figure}[h]
  \centering
  \includegraphics[width=1.0\textwidth]{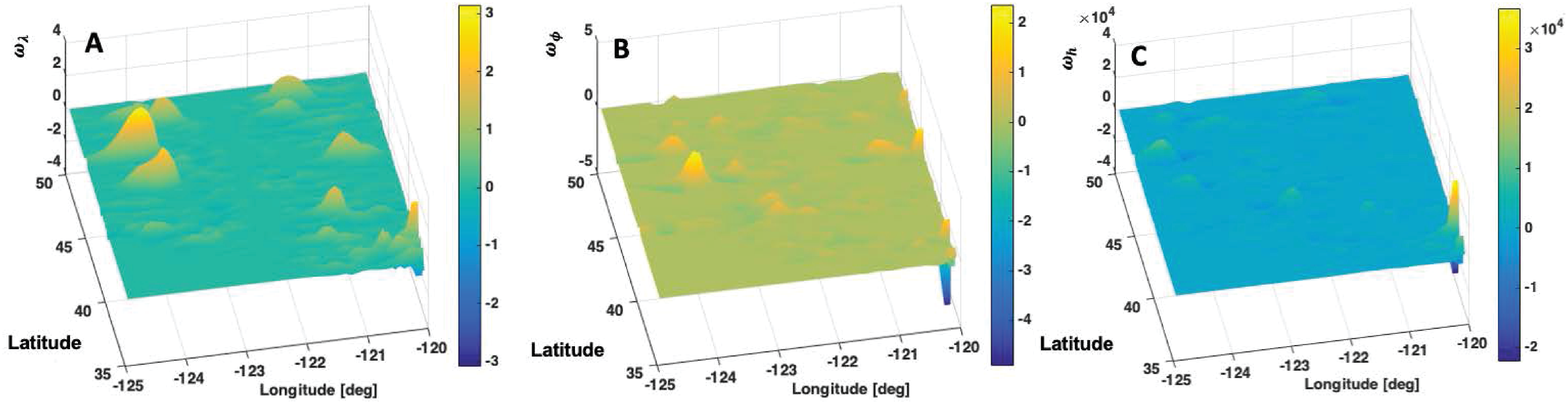}
  \caption{\textbf{Vorticity of the released energy flow at depth = 2.5 km generated from the past 10 years data (epochs from 10355 through 10473).} \textbf{(A-C)} Three components of the vorticity calculated by Eq. (\ref{eq:Vorticity}). }
  \label{fig:PT1_z2_5km_Vorticity}
\end{figure}
%
%
%
\begin{figure}[h]
  \centering
  \includegraphics[width=1.0\textwidth]{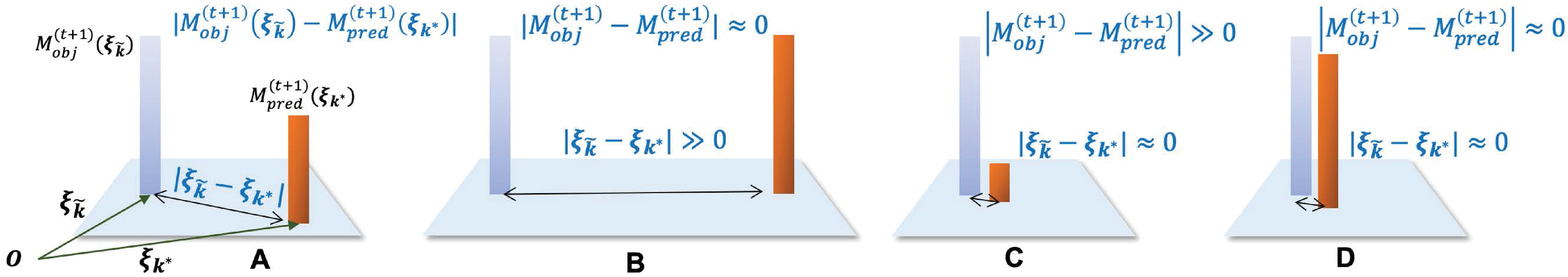}
  \caption{\textbf{Illustration of Comprehensive Error Calculation. (A)} Definition of $M_{obj}$ and $M_{pred}$, the observed and predicted moment magnitudes, respectively, which are closest each other. Examples of poor predictions due to the location error \textbf{(B)} or due to the the magnitude error \textbf{(C)}. Good prediction of both magnitude and location \textbf{(D)}.}
  \label{fig:Mag_Dist_Error}
\end{figure}
\begin{table}[h]
  \caption{Identified free parameters of the best-so-far link functions}
  \label{table:final_free_parameters_LFs}
  \centering
\makebox[1 \textwidth][c]{   
  \begin{tabular}{llll}
    \hline
    LF & $(L_j, T_k)$ & $\textbf{a} = \{a_1, a_2\}$: Exponential LF & Related Rule\\
    \hline
    1  & (10, 3) & $\{1.74118,  0.117647\}$ & Released energy ($E_r$) \\
    2  & (10, 6) & $\{2.77647,  8.03922\}$ & $E_r$ \\
    3  & (25, 3) & $\{2.75294,  4.82353\}$ & $E_r$ \\
    4  & (25, 6) & $\{0.635294,  10.0\}$ & $E_r$ \\    
    \hline
        & Physics & $\textbf{a} = \{a_1, a_2, a_3, a_4, a_5\}$: CRS LF & $\textbf{x}^{*}=\{x_1^*, x_2^*, x_3^*\}$: knots\\
    \hline
    5  & $E_r$ & $\{-0.94902, 1.98431, 1.12157, -0.0705882, 0.980392\}$  & $\{0.169935, 0.624837, 0.682353\}$     \\
    6  & Power & $\{-1.8902, 0.462745, -1.02745, -0.305882, -1.87451\}$  & $\{0.215686, 0.424837, 0.891503\}$     \\
    7  & Vorticity & $\{0.14902, -1.81176, 1.60784, 1.21569, 0.886275\}$  & $\{0.188235, 0.456209, 0.722876\}$     \\  
   \hline
  \end{tabular}
}
\end{table}
%
%
\newpage 
\subsection*{Feasibility test result details}
\begin{figure}[h]
  \centering
  \includegraphics[width=1.0\textwidth]{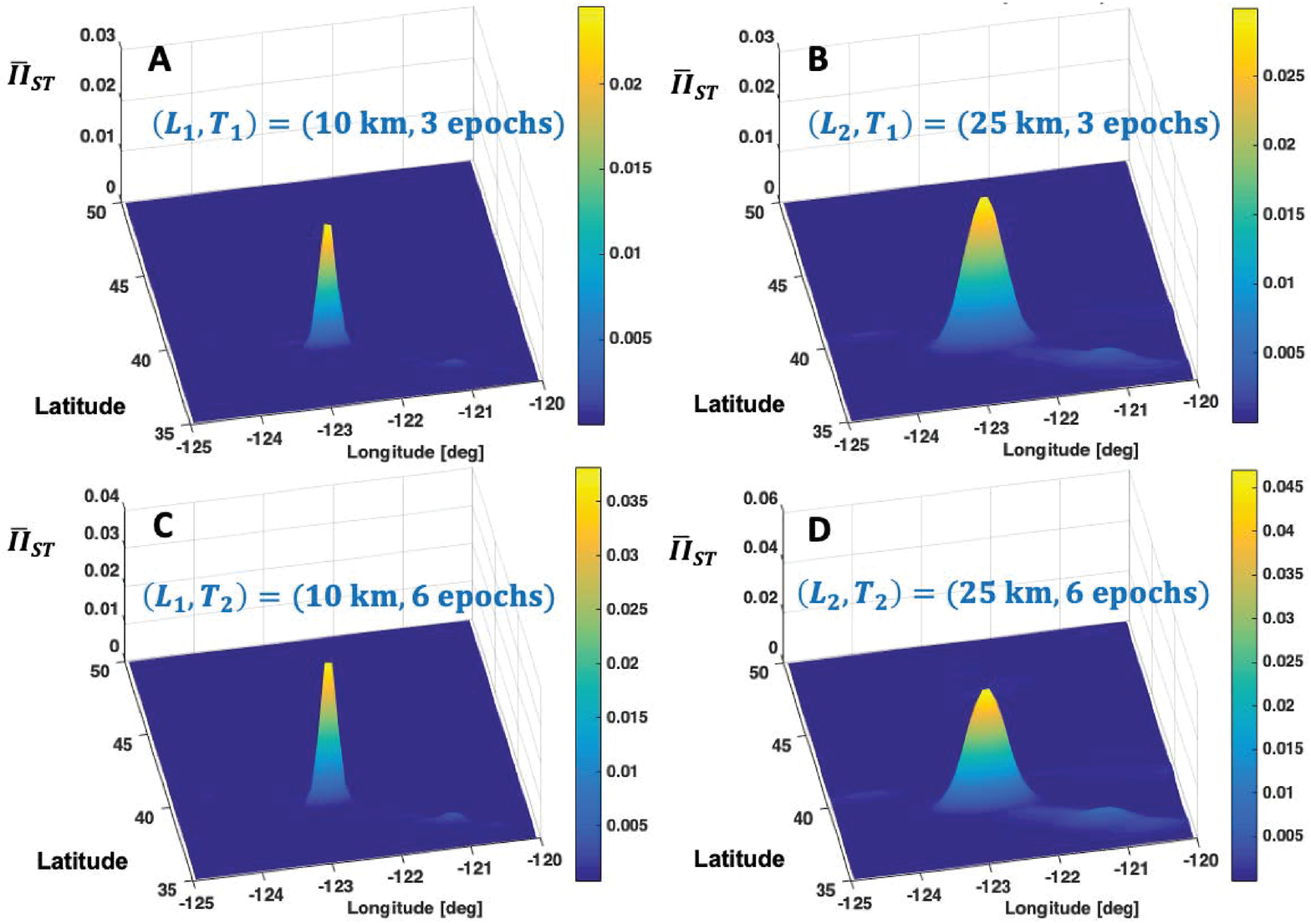}
  \caption{\textbf{Convolved spatio-temporal information index $\overline{II}_{ST}$ at depth = 2.5 km used for the feasibility test with the best-so-far identified rules for the prediction target epoch number 10474 (July, 2019) using past 10 years data (epochs from 10355 through 10473).} \textbf{(A)} At depth = 2.5 km and with the spatial and temporal influence ranges $(L_1, T_1) = $ (10 km, 3 epochs = 3 months). \textbf{(B)} $(L_1, T_2) = $ (10 km, 6 epochs). \textbf{(C)} $(L_2, T_1) = $ (25 km, 3 epochs). \textbf{(D)} $(L_2, T_2) = $ (25 km, 6 epochs).}
  \label{fig:PT1_z2_5km_II_ST}
\end{figure}
\begin{figure}[h]
  \centering
  \includegraphics[width=1.0\textwidth]{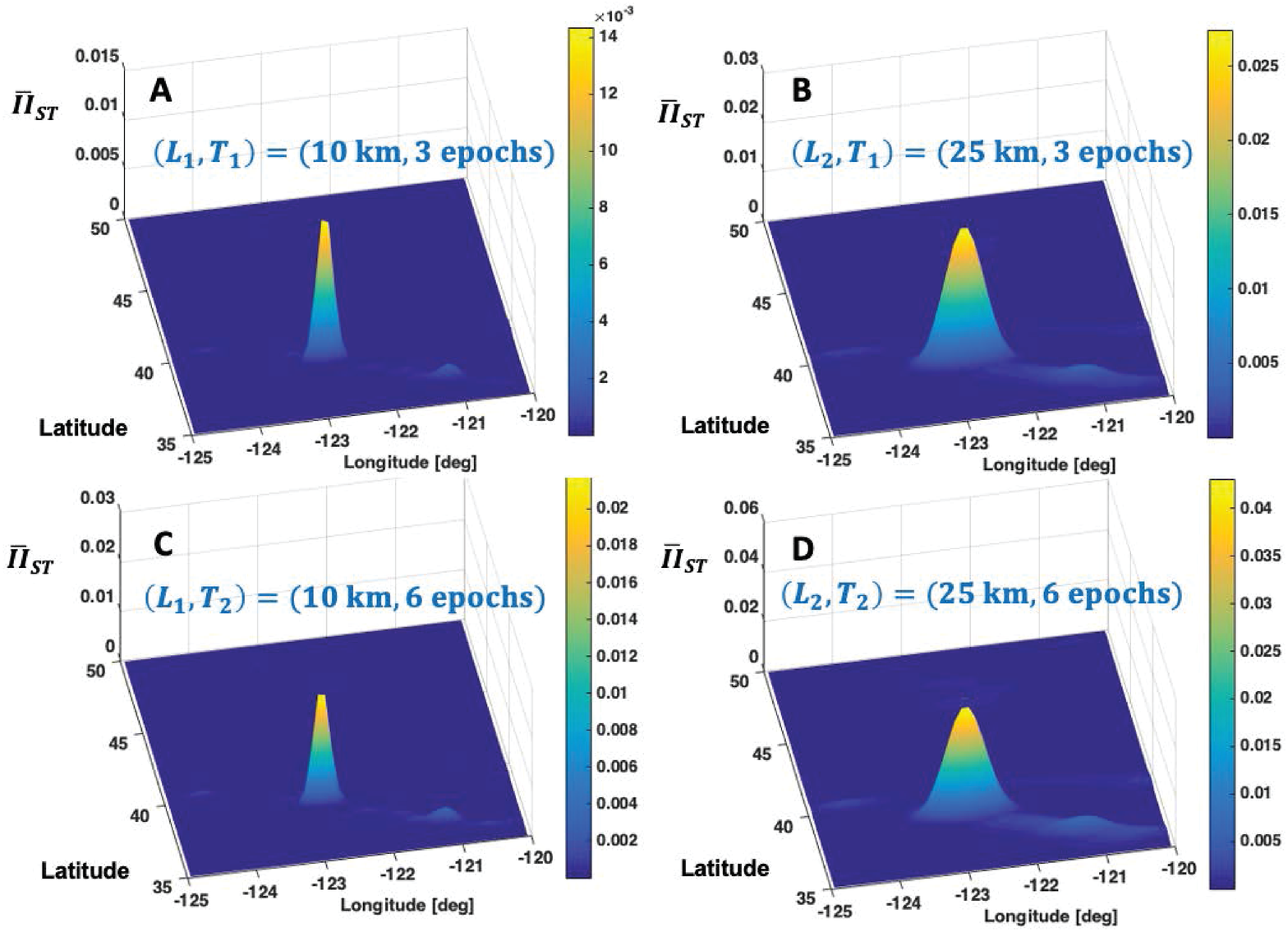}
  \caption{\textbf{Convolved spatio-temporal information index $\overline{II}_{ST}$ at depth = 12.5 km used for the feasibility test with the best-so-far identified rules for the prediction target epoch number 10474 (July, 2019) using past 10 years data (epochs from 10355 through 10473).} \textbf{(A)} At depth = 12.5 km and with the spatial and temporal influence ranges $(L_1, T_1) = $ (10 km, 3 epochs = 3 months). \textbf{(B)} $(L_1, T_2) = $ (10 km, 6 epochs). \textbf{(C)} $(L_2, T_1) = $ (25 km, 3 epochs). \textbf{(D)} $(L_2, T_2) = $ (25 km, 6 epochs).}
  \label{fig:PT1_z12_5km_II_ST}
\end{figure}
\begin{figure}[h]
  \centering
  \includegraphics[width=1.0\textwidth]{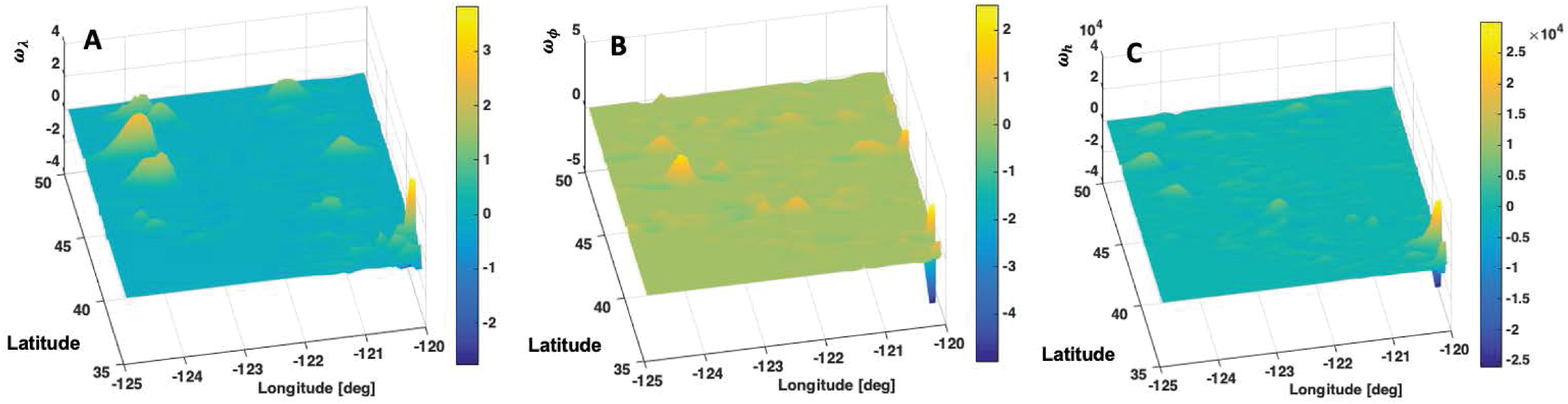}
  \caption{\textbf{Vorticity of the released energy flow at depth = 12.5 km generated from the past 10 years data (epochs from 10355 through 10473).} \textbf{(A-C)} Three components of the vorticity calculated by Eq. (\ref{eq:Vorticity}). }
  \label{fig:PT1_z12_5km_Vorticity}
\end{figure}
The separate feasibility test was conducted on the West-North region of the U.S. as marked by dashed box in Fig. \ref{fig:Prediction_Test1_2_5km}F. The results were predicted by using the identified best-so-far rule of the magnitude prediction rule in Eq. \ref{eq:Magnitude_prediction_CRS_Released_E_Power_Vorticity}. No data exchange across the training and feasibility test regions is allowed. Results were extracted from two separate depths, 2.5 km and 12.5 km, where noticeable earthquake activities are observed (Fig. \ref{fig:Prediction_Test1_2_5km}A and Fig. \ref{fig:Prediction_Test1_12_5km}A). This section presents additional detailed plots of the spatio-temporal convolved II, components of the spatial gradient vector of the released energy, the time derivative of the spatial gradients of the released energy, and components of the pseudo vorticity vector. All values are generated by the identified rules and no intervention was made. 
\newline
\begin{figure}[h]
  \centering
  \includegraphics[width=1.0\textwidth]{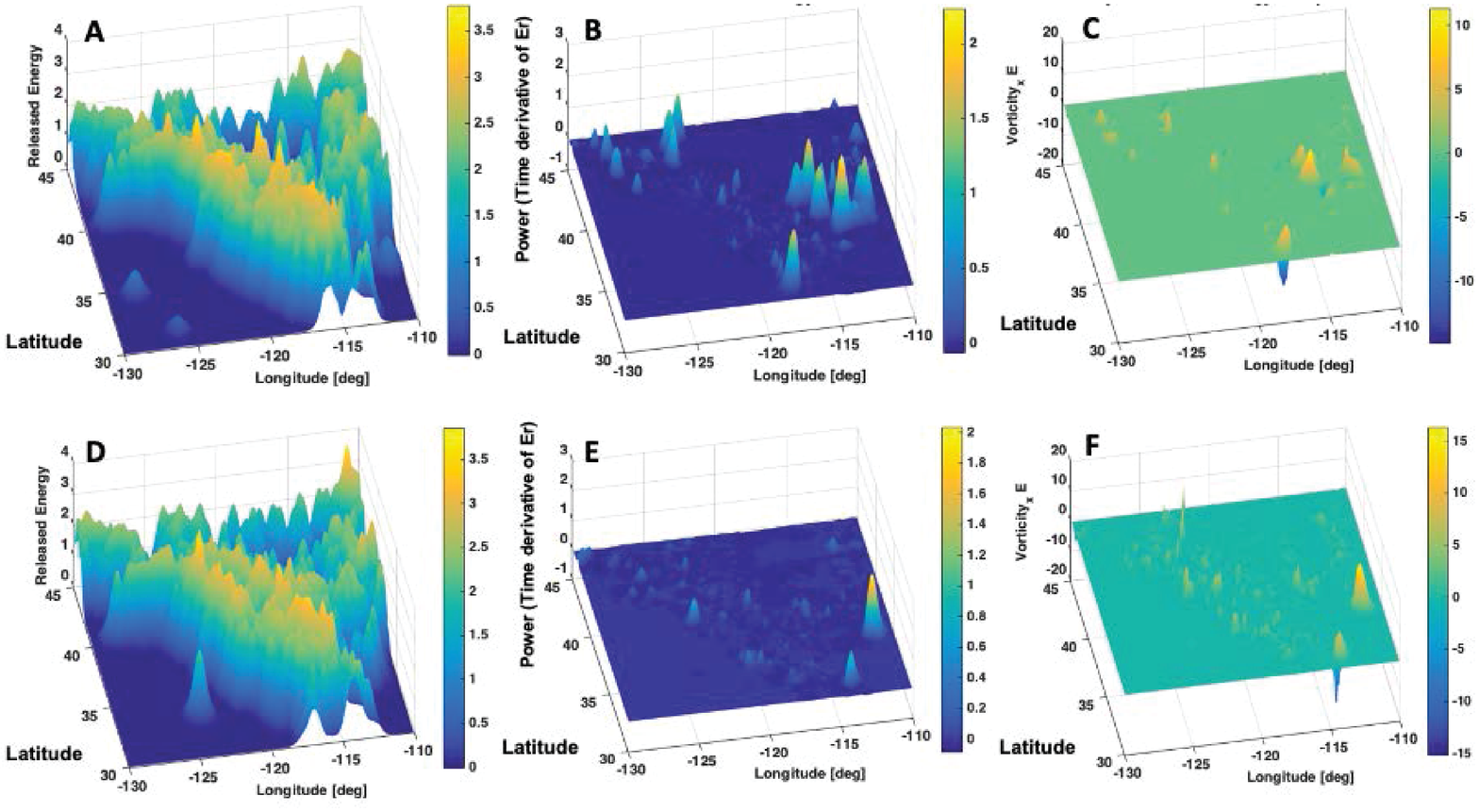}
  \caption{\textbf{Associated results of large scale event prediction test.} \textbf{(A-C)} Released energy, pseudo power, and pseudo vorticity of trained reproduction of Fig. \ref{fig:Test97_PT15}B, respectively. \textbf{(D-F)} Released energy, pseudo power and pseudo vorticity of trained reproduction of Fig. \ref{fig:Test97_PT15}D, respectively.}
  \label{fig:Test97_PT15_pseudo}
\end{figure}
%
\newpage 

\begin{table}
  \caption{Training Setup for the GPRL: Part 1 - Information Index and Learning Control}
  \label{table:train_setup}
  \centering
\makebox[1 \textwidth][c]{  
  \begin{tabular}{lll}
    \hline
    Program syntax & Values & Meaning \\
    \hline
    NUMBER\_TOTAL\_EPOCH\_DATA\_SETS  &  120 & total epoch numbers of training time frame\\  
      &   & the last epoch (10474) is the target \\
      &   & training epochs from 10355 through 10473  \\
    START\_EPOCH\_NUMBER  &  10355 & the first epoch number of the time frame\\
    NUMBER\_INFLUENCE\_RANGES  &  2 & total number of spatial influence ranges $L_j$\\
    INFLUENCE\_RANGES\_VALUES  & (vector)  & values of $L_j$\\    
    10  25  &   & a vector needs to be given below the parameter\\        
    NUMBER\_TIME\_INFLUENCE\_RANGES  &  2 & total number of temporal influence ranges $T_k$\\
    INFLUENCE\_TIME\_RANGES\_VALUES  & (vector)  & values of $T_k$\\
    3  6  &   & a vector needs to be given below the parameter\\
    GLOBAL\_DOMAIN\_RANGES\_VALUES  & (vector) & $(\lambda_{min}, \lambda_{max},\phi_{min},\phi_{max}, h_{min}, h_{max})$\\
    -120 -115 30 40 -10 30  & (vector) & a vector needs to be given below the parameter\\   
    GLOBAL\_DOMAIN\_INCREMENT\_VALUES & (vector) & $\Delta\lambda, \Delta\phi, \Delta h$ of a reference volume \\
    0.1 0.1 5 &   & a vector needs to be given below the parameter\\               
    MIN\_MAGNITUDE\_THRESHOLD & 3.49 & minimum magnitude threshold $M_{thr}$\\
    COEF\_MAGNITUDE\_ERROR & 0.5 & the relative importance factor of magnitude error \\
    & & $a_{M}$ in Eq. (\ref{eq:cost_mag_distance})\\
    VORTICITY\_TYPE & 1 & Pseudo vorticity type in prediction model (Eq. (\ref{eq:Magnitude_prediction_CRS_Released_E_Power_Vorticity}))\\
    & & 1: $\omega_\lambda$; 2: $\omega_\phi$; 3: $\omega_h$; 4: L2 norm of ($\omega_\lambda$, $\omega_\phi$)\\
    VORTICITY\_SCALING\_FACTOR & 7.3890560989 & scaling factor in sigmoid function of vorticity\\    
    & & e.g. $e^2$ = 7.3890560989 for $\omega_\lambda$\\
    R\_MAX\_ERROR & 200 & the maximum range in location error [km] \\
    & & $r_{max}$ in Eq. (\ref{eq:cost_mag_distance})\\    
    COEF\_FALSE\_ALARM\_ERROR & 0.1 & the relative importance of the false alarm error\\
    & & $a_{cnt}$ in Eq. (\ref{eq:cost_type4})\\        
    PREDICTION\_TEST\_UNKNOWN\_TARGET  &  0 & 0: Training; 1: feasibility test    \\  
    BAYESIAN\_UPDATE  &  0 & 1: Perform Bayesian update (B.U.) ; 0: No B.U.    \\      
    PRIOR\_BEST  &  0 & 1: Use the prior-best generation \\
    & & as the initial generation ; 0: Random initialization    \\      
   \hline
  \end{tabular}
}
\end{table}
\begin{table}
  \caption{Training Setup for the GPRL: Part 2 - Evolutionary Algorithm Related}
  \label{table:train_setup2}
\makebox[1 \textwidth][c]{  
  \begin{tabular}{lll}
    \hline
    Program syntax & Values & Meaning \\
    \hline
    NUMBER\_ORGANISMS  &  100000 & total organisms per generation      \\
    N\_ALLELE  &  4 & total alleles per gene     \\
    MUTATION\_RATE  &  0.005 & mutation rate during cross-over and spawning     \\
    MAXIMUM\_GENERATION  &  30 & total generations for evolution     \\
    N\_GENE & 4 & the number of genes per real value \\
    N\_REAL\_VARIABLE & 144 & total number of real-valued free parameters \\
    & &  meaningful up to 56th real variable while others are dummy  \\
    MIN\_MAX\_VALUES & (vector) & min-max range of each free parameter\\
1	0	3	& 		&	$a_1$ (Eq. \ref{eq:multiple_link_exponential}) of the exponential LF number	1	\\
2	0	10	& 		&	$a_2$ (Eq. \ref{eq:multiple_link_exponential}) of the exponential LF number	1	\\
3	-1	1	& 		&	dummy of LF number	1	*\\
4	-1	1	& 		&	dummy of LF number	1	*\\
5	-1	1	& 		&	dummy of LF number	1	*\\
6	0	0.333333333	& 		&	dummy of LF number	1	*\\
7	0.333333333	0.666666667	& 		&	dummy of LF number	1	*\\
8	0.666666667	1	& 		&	dummy of LF number	1	*\\
... & & similar inputs for the exponential LFs number 2 through 4\\
33	-2.0       2.0		& 		&	$a_1$ (Eq. \ref{eq:cubic_spline}) of CRS LF number	5	\\
34	-2.0       2.0		& 		&	$a_2$ of CRS LF number	5	\\
35	-2.0       2.0		& 		&	$a_3$ of CRS LF number	5	\\
36	-2.0       2.0		& 		&	$a_4$ of CRS LF number	5	\\
37	-2.0       2.0		& 		&	$a_5$ of CRS LF number	5	\\
38	0	0.333333333	& 		&	$x_1^*$ (Eq. \ref{eq:cubic_spline}) of CRS LF number	5	\\
39	0.333333333	0.666666667	& 		&	$x_2^*$ of CRS LF number	5	\\
40	0.666666667	1	& 		&	$x_3^*$ of CRS LF number	5	\\
... & & similar inputs for the CRS LFs number 6 and 7\\  
& &  meaningful up to 56th row while others are dummy  \\
   \hline
  \end{tabular}
}
\textbf{* Note: }Each LF is assigned with at most 8 real-valued free parameters for consistency in input. The exponential LF requires only 2 free parameters and thus the remaining 5 values are dummy with no meaning.   
\end{table}
\begin{table}
  \caption{Feasibility Test Setup of GPRL: Part 1 - Information Index and Learning Control}
  \label{table:feasibility_test_setup}
  \centering
\makebox[1 \textwidth][c]{  
  \begin{tabular}{lll}
    \hline
    Program syntax & Values & Meaning \\
    \hline
    NUMBER\_TOTAL\_EPOCH\_DATA\_SETS  &  120 & total epoch numbers of training time frame\\  
      &   & the last epoch (10474) is the target \\
      &   & training epochs from 10355 through 10473  \\
    START\_EPOCH\_NUMBER  &  10355 & the first epoch number of the time frame\\
    NUMBER\_INFLUENCE\_RANGES  &  2 & total number of spatial influence ranges $L_j$\\
    INFLUENCE\_RANGES\_VALUES  & (vector)  & values of $L_j$\\    
    10  25  &   & a vector needs to be given below the parameter\\        
    NUMBER\_TIME\_INFLUENCE\_RANGES  &  2 & total number of temporal influence ranges $T_k$\\
    INFLUENCE\_TIME\_RANGES\_VALUES  & (vector)  & values of $T_k$\\
    3  6  &   & a vector needs to be given below the parameter\\
    GLOBAL\_DOMAIN\_RANGES\_VALUES  & (vector) & $(\lambda_{min}, \lambda_{max},\phi_{min},\phi_{max}, h_{min}, h_{max})$\\
    -125 -120 35 50 -10 30  &   & a vector needs to be given below the parameter\\   
    GLOBAL\_DOMAIN\_INCREMENT\_VALUES & (vector) & $\Delta\lambda, \Delta\phi, \Delta h$ of a reference volume \\
    0.1 0.1 5 &   & a vector needs to be given below the parameter\\               
    MIN\_MAGNITUDE\_THRESHOLD & 3.49 & minimum magnitude threshold $M_{thr}$\\
    COEF\_MAGNITUDE\_ERROR & 0.5 & the relative importance factor of magnitude error \\
    & & $a_{M}$ in Eq. (\ref{eq:cost_mag_distance})\\
VORTICITY\_TYPE & 1 & Pseudo vorticity type in prediction model (Eq. (\ref{eq:Magnitude_prediction_CRS_Released_E_Power_Vorticity}))\\
& & 1: $\omega_\lambda$; 2: $\omega_\phi$; 3: $\omega_h$; 4: L2 norm of ($\omega_\lambda$, $\omega_\phi$)\\
    VORTICITY\_SCALING\_FACTOR & 7.3890560989 & scaling factor in sigmoid function of vorticity\\    
    & & e.g. $e^2$ = 7.3890560989 for $\omega_\lambda$\\    
    R\_MAX\_ERROR & 200 & the maximum range in location error [km] \\
    & & $r_{max}$ in Eq. (\ref{eq:cost_mag_distance})\\    
    COEF\_FALSE\_ALARM\_ERROR & 0.1 & the relative importance of the false alarm error\\
    & & $a_{cnt}$ in Eq. (\ref{eq:cost_type4})\\        
    PREDICTION\_TEST\_UNKNOWN\_TARGET  &  1 & 0: Training; 1: feasibility test    \\  
    BAYESIAN\_UPDATE  &  1 & 1: Perform Bayesian update (B.U.) ; 0: No B.U.    \\      
    PRIOR\_BEST  &  1 & 1: Use the prior-best generation \\
    & & as the initial generation ; 0: Random initialization    \\      
   \hline
  \end{tabular}
}
\end{table}
\begin{table}
  \caption{Feasibility Setup of GPRL: Part 2 - Evolutionary Algorithm Related}
  \label{table:feasibility_test_setup2}
\makebox[1 \textwidth][c]{  
  \begin{tabular}{lll}
    \hline
    Program syntax & Values & Meaning \\
    \hline
    NUMBER\_ORGANISMS  &  5 & Top 5 prior best organisms for feasibility test      \\
    N\_ALLELE  &  4 & total alleles per gene     \\
    MUTATION\_RATE  &  0.005 & mutation rate during cross-over and spawning     \\
    MAXIMUM\_GENERATION  &  1 & One generation for feasibility test     \\
    N\_GENE & 4 & the number of genes per real value \\
    N\_REAL\_VARIABLE & 144 & total number of real-valued free parameters \\
    & &  meaningful up to 56th real variable while others are dummy  \\
    MIN\_MAX\_VALUES & (vector) & min-max range of each free parameter\\
1	0	3	& 		&	$a_1$ (Eq. \ref{eq:multiple_link_exponential}) of the exponential LF number	1	\\
2	0	10	& 		&	$a_2$ (Eq. \ref{eq:multiple_link_exponential}) of the exponential LF number	1	\\
3	-1	1	& 		&	dummy of LF number	1	*\\
4	-1	1	& 		&	dummy of LF number	1	*\\
5	-1	1	& 		&	dummy of LF number	1	*\\
6	0	0.333333333	& 		&	dummy of LF number	1	*\\
7	0.333333333	0.666666667	& 		&	dummy of LF number	1	*\\
8	0.666666667	1	& 		&	dummy of LF number	1	*\\
... & & similar inputs for the exponential LFs number 2 through 4\\
33	-2.0       2.0		& 		&	$a_1$ (Eq. \ref{eq:cubic_spline}) of CRS LF number	5	\\
34	-2.0       2.0		& 		&	$a_2$ of CRS LF number	5	\\
35	-2.0       2.0		& 		&	$a_3$ of CRS LF number	5	\\
36	-2.0       2.0		& 		&	$a_4$ of CRS LF number	5	\\
37	-2.0       2.0		& 		&	$a_5$ of CRS LF number	5	\\
38	0	0.333333333	& 		&	$x_1^*$ (Eq. \ref{eq:cubic_spline}) of CRS LF number	5	\\
39	0.333333333	0.666666667	& 		&	$x_2^*$ of CRS LF number	5	\\
40	0.666666667	1	& 		&	$x_3^*$ of CRS LF number	5	\\
... & & similar inputs for the CRS LFs number 6 and 7\\  
& &  meaningful up to 56th row while others are dummy  \\
   \hline
  \end{tabular}
}
\textbf{* Note: }Each LF is assigned with at most 8 real-valued free parameters for consistency in input. The exponential LF requires only 2 free parameters and thus the remaining 5 values are dummy with no meaning.   
\end{table}
%
\subsection*{Bayesian update and evolutionary algorithm}
Aiming at no distributional assumptions about the priors/posteriors as well as pursuing smooth evolution, this study adopts the combination of Bayesian update and a modified genetic algorithm \cite{Cho:2019}. The key evolutionary algorithm involves the preparation of initial generation, organism-wise evaluation of fitness score, and fitness-based spawning of the next generation. The prior best physical rules can be smoothly inherited by the Bayesian update-based fitness proportionate probability (FPP) rule. To accelerate the evolution speed of the modified genetic algorithm, an individual variable-wise gene cross-over scheme has been used, and the changing search range scheme is used in an iterative manner for better performance as successfully done in \cite{Cho:2019}. Since an individual $s$ realizes a candidate of $\bm{\uptheta}$, the all free parameters in current generation $S$, the raw cost of an individual $s$, termed as $\mathcal{J}(s)$, is calculated by a number of types. In the type definition, $M_{obs}^{(t+1)}(\bm{\xi}_j)$ means the observed maximum moment magnitude in the $j$th reference volume at epoch $(t+1)$ that is regarded as the true (measured) physical response. $M_{thr}$ denotes the user-defined moment magnitude threshold. 
Then, following typical genetic algorithm procedure \cite{Koza:1992, Cho_et_al:2018, Cho:2019} the normalized fitness score $\mathcal{F}$ of an individual is calculated by 
\begin{equation}\label{eq:fitness}
\mathcal{F}(s) = \frac{ (1+ \mathcal{J}(s))^{-1}}{ \sum_{\forall s \in S} [(1+ \mathcal{J}(s))^{-1}]}
\end{equation}
where $s$ denotes an individual in the entire generation $S$. 
Learning a hidden physical rule is not a one-time task, rather a continuous activity. As diverse new experimental data become available, the physical rule learner must embrace all the previous knowledge and learn new information. To seamlessly realize this continuous learning, this study infused the Bayesian update scheme into the evolutionary algorithm's FPP rule. 
Suppose we have the best-so-far generation, denoted as $S^*$ and its associated fitness scores, $\mathcal{F}^{*}(s)$, $s \in S^*$. According to the FPP rule, the probability of selecting an $\bm{\uptheta}$ for next parent is given by $\textrm{Prob}(\bm{\uptheta}) \propto \mathcal{F}(s), s \in S^*$. Thus, $\mathcal{F}^{*}(s)$ is regarded as a prior PDF of parameters $\bm{\uptheta}=\{\textbf{a}, \textbf{x}^* \}$, i.e. $\pi_{prior}(\bm{\uptheta})$ in the typical Bayesian formalism. For initialization of $\pi_{prior}(\bm{\uptheta})$, this study intentionally departs from fully random initialization to investigate positive evolution trends without special initialization assumption. Thus, this framework is purely data-driven, requiring no distributional assumptions about priors and posteriors. 
For the posterior distribution, we adopted the following two-stage procedure. \\Suppose that we have the prior best LFs and their $S^*$ and that new experimental data become available. At the first learning generation with the new data, we can calculate the first fitness scores $\mathcal{F}(s; S^*)$ by applying the prior $S^*$ and LFs to the new experiment. After the first generation, we can estimate the Bayesian fitness score (denoted as $\mathcal{F}_B$) as:
\begin{equation}\label{eq:F_B}
\mathcal{F}_B(s) = \frac{1}{\kappa} 
    \frac{\mathcal{F}(s;S^{*}) \mathcal{F}^*(s)}{\sum_{\forall s\in {S^*}} \mathcal{F}^*(s)}
\end{equation}
where $\kappa$ is needed for normalizing the Bayesian fitness to unity, which is simply given by  
\begin{equation}\label{eq:kappa}
\kappa =  \sum_{\forall s\in {S^*}} \frac{\mathcal{F}(s;S^{*})\mathcal{F}^{*}(s)}
             {\sum_{\forall s\in {S^*}} \mathcal{F}^*(s)}
\end{equation}
Then, from the second learning generation of the new experiment, the probability of selecting two parents is proportional to the Bayesian fitness score as 
\begin{equation}\label{eq:parent}
\textrm{Prob}(\textrm{parent}_i | s) \propto \mathcal{F}_B (s),~(i=1,2).
\end{equation}
Once again, an individual $s$ realizes a candidate of $\bm{\uptheta} = (\textbf{a}, \textbf{x}^{*})$ in the new generation $S$, and thus the desired posterior distribution is obtained. In this way, the prior knowledge is smoothly inherited to the new experiment on the framework of evolutionary algorithm, thereby enabling constantly evolving physical rule learning. For allowing for evolving with new data, the previous scores are inherited by the Bayesian score Eq. (\ref{eq:F_B}). Since the adopted evolutionary algorithm remembers prior generation's fitness scores, which offer the probability distribution of free parameters of LFs. As the Bayesian inheritance continues with new experimental data, the probability distribution of LFs will naturally evolve. Thus, the proposed framework can achieve evolving capability with increasing data. In the future, more dedicated investigations should focus on validation of the constantly evolving capability of LFs with sufficient, diverse test data. To some extent, the aforementioned combination of Bayesian update and evolutionary algorithm can be viewed as a log-likelihood maximization as explained in \cite{Cho:2019}. 
%
%

\clearpage

\end{document}